\documentclass[
reprint,
superscriptaddress,
frontmatterverbose,
amsmath,amssymb,
aps,
]{revtex4-2}

\usepackage{graphicx}
\usepackage{dcolumn}
\usepackage{bm}
\usepackage{mathrsfs}

\usepackage{color}
\usepackage[version=4]{mhchem}

\renewcommand\vec{\bm}

\newcommand{\bk}{\vec{k}}
\newcommand{\br}{\vec{r}}

\usepackage{bbm}
\newcommand{\ii}{{\mathrm{i}\,}}
\usepackage{setspace}
\usepackage{lineno}

\begin{document}


\title{Neural network representation of electronic structure \\from \textit{ab initio} molecular dynamics}

\author{Qiangqiang Gu}
\affiliation{International Center for Quantum Materials, School of Physics, Peking
University, Beijing 100871, China}

\author{Linfeng Zhang}
\affiliation{Department of Mathematics and 
Program in Applied and Computational Mathematics, Princeton University, Princeton, NJ 08544, USA}

\author{Ji Feng}
\email{jfeng11@pku.edu.cn}
\affiliation{International Center for Quantum Materials, School of Physics, Peking
University, Beijing 100871, China}
\affiliation{Collaborative Innovation Center of Quantum Matter, Beijing 100871,
China}

%

\date{\today}

\begin{abstract}
  Despite their rich information content, electronic structure data amassed at high volumes in \textit{ab initio} molecular dynamics simulations are generally under-utilized. We introduce a transferable high-fidelity neural network representation of such data in the form of tight-binding Hamiltonians for crystalline materials. This predictive representation of \textit{ab initio} electronic structure, combined with machine-learning boosted molecular dynamics, enables efficient and accurate electronic evolution and sampling. When applied to a one-dimension charge-density wave material, carbyne, we are able to compute the spectral function and optical conductivity in the canonical ensemble. The spectral functions evaluated during soliton-antisoliton pair annihilation process reveal significant renormalization of low-energy edge modes due to retarded electron-lattice coupling beyond the Born-Oppenheimer limit. The availability of an efficient and reusable surrogate model for the electronic structure dynamical system will enable calculating many interesting physical properties, paving way to previously inaccessible or challenging avenues in materials modeling.

\end{abstract}

\maketitle

\section{Introduction} 
\textit{Ab initio} molecular dynamics (AIMD) is an atomistic simulation method based on \textit{ab initio} electronic structures, and a versatile tool for quantitative understanding thermodynamic and kinetic behaviors of wide range of matter, including molecules, liquids, and solids \cite{Parrinello1985,Hutter2009}.
In an AIMD simulation, a huge number of atomic configurations have to be sampled for meaningful statistics, during which the Born-Oppenheimer (BO) electronic structure as a function of atomic trajectory is evaluated at \textit{ab initio} level. Although a number of electronic properties can be computed on-the-fly, the electronic structure data, generated at high volumes and significant computation costs, are generally under-utilized, despite exceedingly rich information in these data. This is precisely a situation where a concise and predictive representation of the \textit{ab initio} electronic structure data to harvest the electronic information discarded after AIMD is expected to be highly rewarding. This will constitute a reusable surrogate model, which can significantly expedite the sampling of configuration-dependent electronic structure. 

The above problem falls squarely in the realm of deep neural network, which has become a powerful and versatile paradigm for computerized analysis and interpretation of massive data \cite{LeCun2015}. 
Through supervised observation of labeled data, a neural network automatically recognizes patterns and regularities in the data and hence becomes predictive regarding the features it is devised to represent. In turn, the predictive neural network can substitute the originally time-consuming calculation with significant speed-up. 
For example, a neural network representation of the interatomic potential has been developed and can reproduce the atomic trajectory without time-consuming \textit{ab initio} calculations\cite{Behler2007,  Schutt2017, Schutt2018, Linfeng2018, Linfeng_NIPS_2018}.  
In these methods, the total potential energy is decomposed into a sum of atomic energy contributions, which is represented by neural networks through the atom and its neighbors (local chemical environment). However, the electronic information available in an AIMD is entirely lost in this approach. Therefore, a predictive representation of electronic structure to pick up the rich electronic structure information in AIMD simulation is highly desirable, which will enable many previous inaccessible or challenging tasks in modeling the electronic properties of a large and dynamical system. 

Recently, progresses have been made in surveying  electronic properties using machine learning (ML) techniques, focusing exclusively on specific materials properties from structural information, such as the electronic density of states \cite{Mahmoud2020} and charge density \cite{Chandrasekaran2019}, or focusing on electronic Hamiltonians for small molecular systems \cite{Schuett2019}. 
A general ML approach that predicts a microscopic electronic Hamiltonian for crystalline materials from structural information is still lacking, which shall enable direct computation of physical properties with the high efficiency offered by ML techniques. In this paper, we present a neural network architecture to efficiently predict a tight-binding Hamiltonian (dubbed TBworks hereafter) to represent the electronic structure data from AIMD simulations. This architecture will employ a deep feedforward network to process atom trajectory and band structures, to produce a representation that predicts a tight-binding Hamiltonian for a given snapshot of atomic configuration. The main challenge in our method is that the neural network predicted tight-binding Hamiltonian is not uniquely determined by its band structure due to the $U(N)$ gauge freedom, which calls for much effort in designing the neural network architecture and optimization process. It is shown that by the elaborate design of the neural network architecture, the gauge freedom can be reduced. After introducing the architecture and algorithm, the method is tested using a one-dimension charge-density-wave (CDW) material, carbyne. It is shown that the \textit{ab initio} electronic structure can be faithfully reproduced by the predicted Hamiltonian. 

As an example, large-scale molecular dynamics (MD) simulations are performed on carbyne based on the ML force field, during which the TBworks is applied to sample the electronic structures and derived physical properties. 
This enables us to sample the electronic states in the dynamical annihilation of a soliton-antisoliton pair to compute the time-dependent spectral function, beyond the Born-Oppenheimer approximation. Optical conductivity is also obtained by sampling the current-current correlation function in a path-integral molecular dynamics. These results demonstrate that our TBworks, combined with ML force field, realizes a paradigm for expedited electronic structure sampling at \textit{ab initio} accuracy, which can capture the non-adiabatic effects in transient processes by large-scale and long-time simulations.

\section{Neural Network Architecture} 

In an AIMD simulation, a trajectory of ion positions is generated at a discrete set of time points, $\bm{u}_i(t) = \br_i(t) - \bar{\br}_i$, where $\br_i(t)$ and $\bar{\br}_i$ are respectively the instantaneous and the reference ion positions. Periodic boundary conditions are imposed on the simulation box, such that the \textit{ab initio} eigenstates are Bloch functions. An AIMD yields a set of instantaneous electronic eigenstates for an ionic configuration $\br= \{\br_i\}$, 
with corresponding energy spectrum (bands) $\hat{\varepsilon}(\br) = \{\hat{\varepsilon}_{n\bk}\}$. 
The aim is then to use $\br$ and $\hat{\varepsilon}(\br)$ to train a neural network as a surrogate model for the \textit{ab initio} electronic structure, through a tight-binding electronic Hamiltonian as a function of $\br$.
\begin{eqnarray}
	H(\br) = \sum_{i\neq j} V_{ij}(\br) c_{i}^\dagger c_j + \sum_{i} V_{ii}(\br) c_{i}^\dagger c_i + \text{H.c.},
\end{eqnarray}
where $V_{ij}(\br)$ $(i\neq j)$ is a hopping matrix element and $V_{ii}(\br)$ is  the on-site energy. Assuming that there is only a single orbital on each ion for notational brevity, $i$ and $j$ are also site indices.

\begin{figure*}[htbp]
\centering
  \includegraphics[width=120 mm]{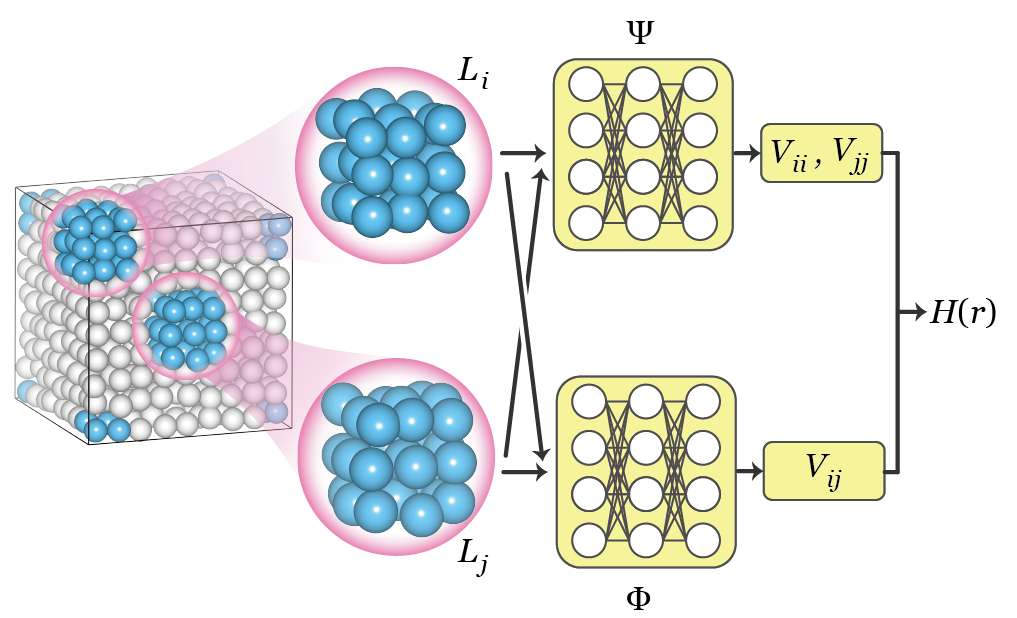}
  \protect\caption{(color online) Schematic of the local chemical environment construction and the TBworks neural network structure. Ions in local environments $L_i$ and $L_j$ are depicted as blue balls enclosed by translucent spheres. $\Psi$ and $\Phi$ are multilayer fully connected neural networks with a hyperbolic tangent activation (open circles). }
	\label{fig:1}
\end{figure*}

Providing the existence of Wannier functions \cite{Brouder07, Cloizeaux1964, nenciu1983}, it suffices to keep hopping matrix elements between pairs of orbitals within a finite range. Therefore, we introduce a \emph{local chemical environment} for each site, as shown in Fig. \ref{fig:1}. 
For site $i$, the ions lying within a sphere of radius $R_{\mathrm{cut}}$ centered at $\br_i$ form its local chemical environment $L_i=\left\{j \ | \ | \br_j-\br_i | < R_{\mathrm{cut}}\right\}$, which assumes non-negligible influence on the Wannier functions on site $i$. Numerically, the local environment $L_i$ is represented by the set $D_i = \left\{ \br_{ji} | j \in L_i  \right\}$, where $\br_{ji} = \br_j - \br_i$. When fed to neural network, the elements in set $D_i$ are sorted by distances $|\br_{ji} |$ in ascending order.
Therefore, the hopping matrix element $V_{ij}$ will depend on $\bar{\br}_{ji}$, $\bm{u}_i$, $\bm{u}_j$, $D_i$, and $D_j$, where $\bar{\br}_{ji} = \bar{\br}_{j}-\bar{\br}_{i}$. The on-site energy $V_{ii}$ can be fully determined by $D_i$.
A na\"ive neural network for our purpose is composed of two maps (see Fig.~\ref{fig:1}): $V_{ij}=\Phi(X;\theta)$ and
$V_{ii} = \Psi(Y;\xi)$, where $X=\left(\bar{\br}_{ij}, \bm{u}_i, \bm{u}_j, D_i, D_j\right)$, $Y=D_i$. $\theta$ and $\xi$ are trainable parameters of the neural network. The tight-binding Hamiltonian $H(\bar{\br})$ for reference configuration (usually an equilibrium crystal structure) is easily accessible with standard Wannierization method \cite{Marzari12}, which is used to calibrate the neural network in a refined architecture,
\begin{eqnarray}
\begin{aligned}
	V_{ij} &= \bar V_{ij} + \Phi(X;\theta) - \Phi(\bar{X};\theta),\\
	V_{ii} &= \bar{V}_{ii} + \Psi(Y;\xi) - \Psi(\bar{Y};\xi).
\end{aligned}
\end{eqnarray}
Obviously, when $\br=\bar{\br}$, $\Phi(X;\theta) - \Phi(\bar{X};\theta)$ and $\Psi(Y;\xi) - \Psi(\bar{Y};\xi)$ equal zero for the neural network to produce exactly $H(\bar{\br})$. Although we only show the neural network maps for the case of single orbital on each ion for brevity, generalization to multi-orbital case is straightforward by allocating individual maps for hoppings between different orbitals.

The maps $\Phi$ and $\Psi$ are fulfilled through deep feedforward neural networks with $n$ hidden layers neurons, as shown in Fig.~\ref{fig:1}. Take the map $\Phi$ as an example, 
$
	\Phi = \Phi_n \circ \Phi_{n-1} \circ \ldots  \circ \Phi_1.
$ 
The $\mu$-th  hidden layer, $\Phi_\mu$ takes the output vector of the preceding layer $x_{\mu-1}$, and generates a new vector through the component-wise  hyperbolic tangent activation,
$
	x_\mu = \tanh(\\w_\mu x_{\mu-1} +b_\mu),
$
with $x_0= X$, and $V_{ij}= x_n$. Here the matrix $w_\mu$ reweights $x_{\mu-1}$ and the vector $b_\mu$ provides a bias, both of which are adjustable parameters and determined via the Adam stochastic gradient descent method \cite{Adam2014} employed to minimize the loss function during a back-propagation process. The loss function is the root-mean-square (rms) deviation  between energy spectra $\varepsilon (\br)$ computed from predicted $H(\br)$ and training labels from the \textit{ab initio} $\hat{\varepsilon}(\br)$.
In principle, $H(\br)$ is not uniquely determined by energy spectra $\hat{\varepsilon}(\br)$ alone due to the existence of gauge freedom. If a different gauge is chosen at different $\br$, $H(\br)$ as a function of $\br$ may not necessarily be smooth and continuous. Because of the discontinuity, the predicted $H(\br)$ may not interpolate well for unseen configurations, not to mention extrapolations. Fortunately, the continuity of $H(\br)$ is fulfilled automatically in our neural network implementation, since the relationships between $\br$ and $H(\br)$ are represented by $\Phi$ and $\Psi$, which are smooth and continuous composite maps. A simple test of whether our method is plagued by the gauge uncertainty is its ability to generalize.  As will be shown next, our TBworks generalizes accurately on all the independent testing data sets. 

A couple of remarks are warranted at this point, on how the method outlined above is different from some of the related existing techniques. First, our TBworks and ML force fields  methods \cite{Behler2007,  Schutt2017, Schutt2018, Linfeng2018, Linfeng_NIPS_2018} offer completely different surrogate models for AIMD simulations. Whereas ML force fields predict the interactions potential energies and forces between ions, TBworks predicts electronic Hamiltonians for given ion configurations. Therefore, although both methods are designed to remove the need for time-consuming AIMD, ML force fields are fashioned to accelerate molecular dynamics, while our TBworks will enable fast sampling of electronic structure, for instance, in a ML-force-field accelerated MD. Second, TBworks is not the semi-empirical density-functional tight-binding method \cite{porezagdftb1995,Elstnerdftb1998}. The latter is designed to speed up a density-functional theory calculation with the help of an approximate tight-binding Hamiltonian. Our TBworks, on other hand, produces a very accurate representation of Kohn-Sham electronic structure itself.

To sum up, we have outlined the essential idea of TBworks in this section. In the following sections, we will present an exemplary application of the TBworks on a simple yet nontrivial material, to demonstrate that TBworks can accurately and efficiently predict electronic structure of a dynamical system to facilitate the computation of physical properties. With this example, more details of the TBworks approach, including data preparation, network training energy windowing, electronic structure sampling, generalizability and transferability, and the interfacing with force fields MD simulation, will be clarified.

\section{Accuracy and transferability}
\begin{figure}[htbp]
	\centering
		\includegraphics[width=70 mm]{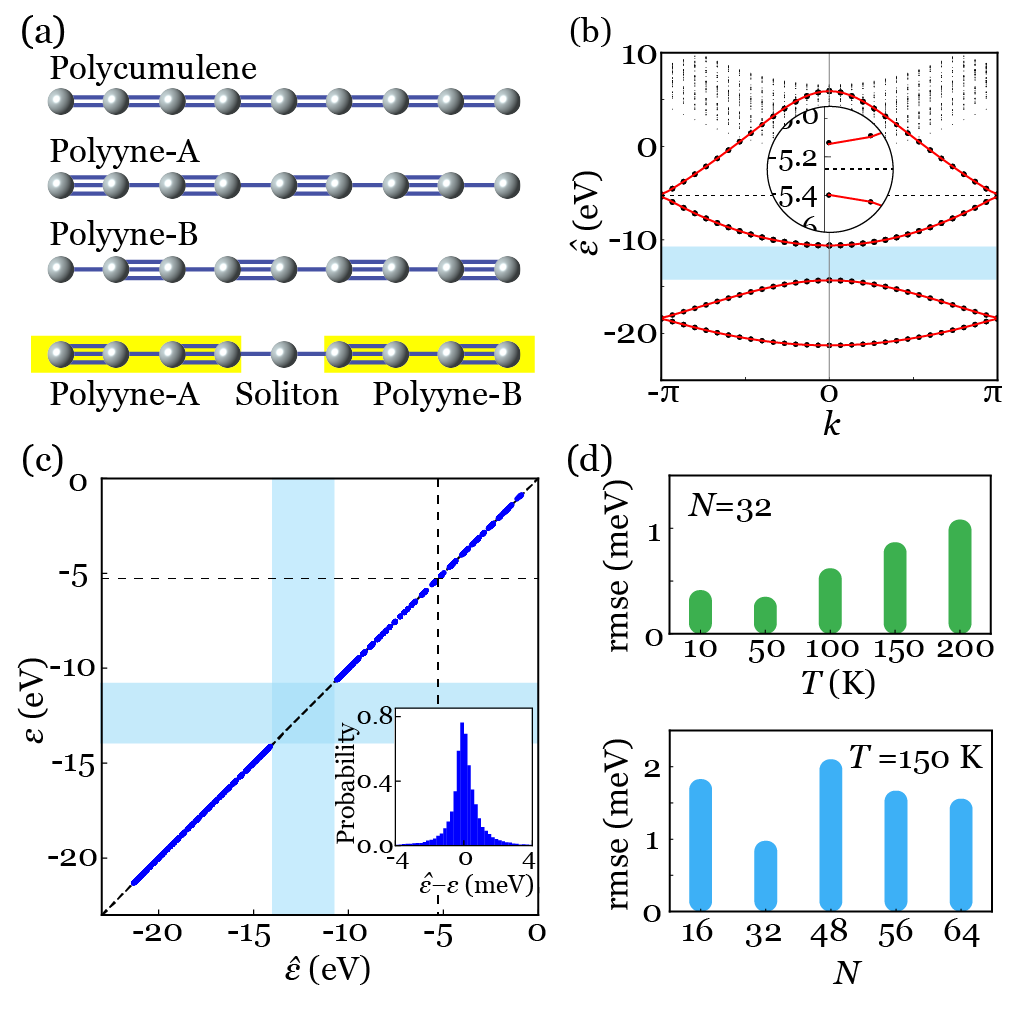}
		\protect\caption{(color online) Performance of neural network representation of the electronic structure of carbyne. (a) Structure of carbyne, including (from top to bottom) polycumulene, two polyynes, and a chain with a kink excitation. (b) TBworks (red line) and \textit{ab initio} (black dots) band structure for polyyne. Blue shaped region indicates the $s$-$p$ separation, and the black dashed line indicates the CDW gap (a blow-up view in the inset). (c) TBworks predicted vs \textit{ab initio} eigenvalues for unseen snapshots with different simulation box sizes and temperatures, $T=10, 50, 100, 150, 200$ K and $N=16, 32, 48, 56, 64$. Inset: error distribution. (d) Testing root-mean-square error (rmse) for unseen snapshots with a simulation box size $N=32$ at different temperatures (upper panel) and snapshots at $T=150$~K with different simulation box sizes (lower panel).}
	\label{fig:2}
\end{figure}

We now proceed to evaluate the performance of our TBworks algorithm by learning and then predicting AIMD eigenvalues for carbyne \cite{Heimann1983, Chuanhong2009, Cretu2013}. Carbyne is a one-dimensional $sp^1$ hybridized carbon chain, which can undergo an archetypal Peierls' transition from polycumulene (\ce{=C=C=}) with uniform C-C bond lengths, to a CDW phase, polyyne (\ce{-C#C-}) \cite{Kertesz1978,Peierls1996,Su79}. The polyyne  has two degenerate ground state structures interrelated by a shift of the single and triple bonds, as shown in Fig.~\ref{fig:2}(a). 
Polyyne can be characterized by the order parameter 
$u_\alpha=(-1)^\alpha u_0$ $(\alpha=1,2,\cdots,N)$,  
and sign$(u_0)=\pm 1$  corresponds to the two degenerate phases. Remarkably, sign$(u_0)=\pm 1$ can coexist on a single carbyne chain, separated by a domain wall or a kink in $u_\alpha$,  which can lead to solitonic excitations \cite{Su79}. The CDW and solitonic excitations are the manifestations of correlations between electronic and ionic degrees of freedom, which bring about many intriguing physical phenomena such as spontaneous symmetry breaking, polarons and superconductivity, etc. The CDW bond ordering and solitonic excitations make carbyne an interesting and non-trivial material for testing our algorithm. 

The AIMD simulations of carbyne in the canonical ensemble are performed by the Vienna \textit{ab initio} simulation package within the generalized-gradient approximation \cite{Kresse1996,Perdew1981} to generate the training data for TBworks.
Nose-Hoover thermostat \cite{Nose1984,Hoover1985} is used to maintain the temperatures. The cutoff energy is 400 eV for wave functions expansion in the plane-wave basis sets.  The AIMD simulations are performed with the simulation box sizes containing $N=16$, $32$, $48$, $56$ and $64$ carbon atoms and at temperatures $T=10$, $50$, $100$, $150$, and $200$~K. Periodic boundary conditions are applied to all three directions of the simulation box, which is a cuboid $\vec a\times \vec b\times \vec c$, with mutually orthogonal $\vec a$, $\vec b$ and $\vec c$. The carbyne chain is aligned along $\vec c$-direction and $a=b=20$~\AA.  The eigenvalue labels are calculated using $1\times1\times4$ and $1\times1\times2$ $\bk$-mesh for $N=16$ and $N=32$ boxes, while for $N>32$, only the $\Gamma$ point is used. In order to train the TBworks, labeled data composed of 5000 snapshots from one AIMD trajectory with $N=32$ at $T=150$~K are used as training data. The determination of the system size for training data and the performance of TBworks trained by data with $N=48$ are shown in Appendix  section C and D. The generalizability and transferability of TBworks are validated against multiple unlearned data sets (5000 AIMD snapshots each) from multiple $T$ and $N$. We use the term generalizability to denote the ability of TBworks to predict accurately for unseen data for the same chemical compound, under the same conditions (system size $N$ and temperature $T$) as training data. In contrast, by transferability we mean the robustness of TBworks for the same compound but under varied conditions ($N$ or $T$, in this context).

In carbyne, the C-C $\sigma$ bonds are formed by the overlap of $sp^1$ hybridized orbitals while the $\pi$ bonds are formed by the other two half-filled $p$ orbitals perpendicular to the carbon chain. The $s$ and $p$ bands within the energy window (-22 -- 0 eV) are picked out as label data from 5000 AIMD snapshots of $N=32$ at $T=150$~K to train the 3-layer TBworks with the [200, 200, 200] neurons. Fig.~\ref{fig:2}(b) displays a faithful reproduction of the band structure of polyyne by the trained TBworks, including the CDW gap $\Delta\sim$ $0.27$ eV. In Fig.~\ref{fig:2}(c), the TBworks predicted spectra for snapshots unseen in training are plotted against \textit{ab initio} spectra, where samples are drawn from AIMD with multiple temperatures (10 -- 200~K) and a range of simulation box sizes. The testing error (rms deviation) of these predictions is  $\sim 1$ meV and the coefficient of determination $R^2=0.99999994$. The $s$-$p$ separation of $\sim 3.72$~eV and the CDW gap $\sim$ $0.27$~eV are also well reproduced despite their disparate energy scales.  As is also shown in Fig.~\ref{fig:2}(d), the testing errors of the TBworks trained at a single $T$ and $N$ are on the order of $1$~meV, when tested against data from multiple temperatures and simulation box sizes. 

\begin{figure}[ht]
	\centering
		\includegraphics[width=70 mm]{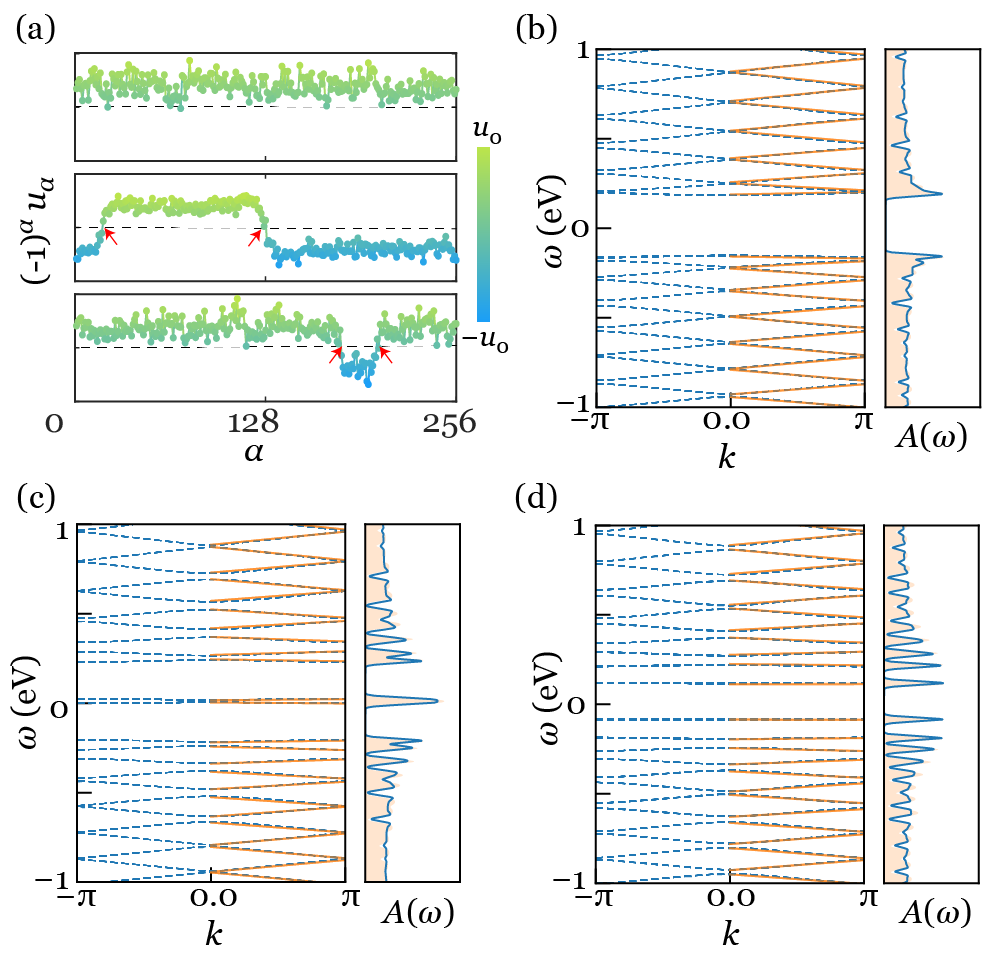}
	\protect\caption{(color online) Comparison between TBworks predicted and \textit{ab initio} calculated electronic structure, for three snapshots in a MD simulation of a 256-site carbyne chain. (a) Order parameters of the three configurations. Top: a polyyne configuration; Middle: an isolated soliton pair; And bottom: a coupled soliton pair. The red arrow point to the soliton locations. (b-d) The TBworks-predicted (blue line) vs \textit{ab initio} (yellow line and zone) band structures and density of states of the three configurations in (b), in the same order.}
	\label{fig:3}
\end{figure} 

Further generalization tests have been performed on a long-chain carbyne with $N=256$, without or with a soliton pair. The order parameter values of several snapshots from the MD trajectory at $T=10$~K based on the ML force fields \cite{Linfeng_NIPS_2018} are shown in Fig.~\ref{fig:3}(a). Three archetypal structures, polyyne (top), isolated soliton/antisoliton (middle) and coupled soliton-antisoliton pair (bottom), are selected for the tests.  The band structures and density of states from TBworks and \textit{ab initio} benchmarks for the three representative configurations are respectively shown in Fig.~\ref{fig:3}(b-d). Evidently, the TBworks faithfully reproduces full energy spectra in the low-energy regime. In particular, the CDW gap in polyyne structure (Fig.~\ref{fig:3}(b)), the zero-energy soliton mode in isolated soliton/antisoliton structure (Fig.~\ref{fig:3} (c)) are accurately reproduced. So are the incipient in-gap modes with a finite splitting when the soliton and antisoliton experiences coupling at short separations, as shown in Fig.~\ref{fig:3}(d).
Therefore, we conclude that the TBworks trained at a single $T$ and $N$ offers a high fidelity representation of the \textit{ab initio} electronic structure, which is also highly transferable to a range of temperatures, system sizes, and especially, structural variety, unseen in the initial learning process.

\begin{figure}[htbp]
	\centering
		\includegraphics[width=70 mm]{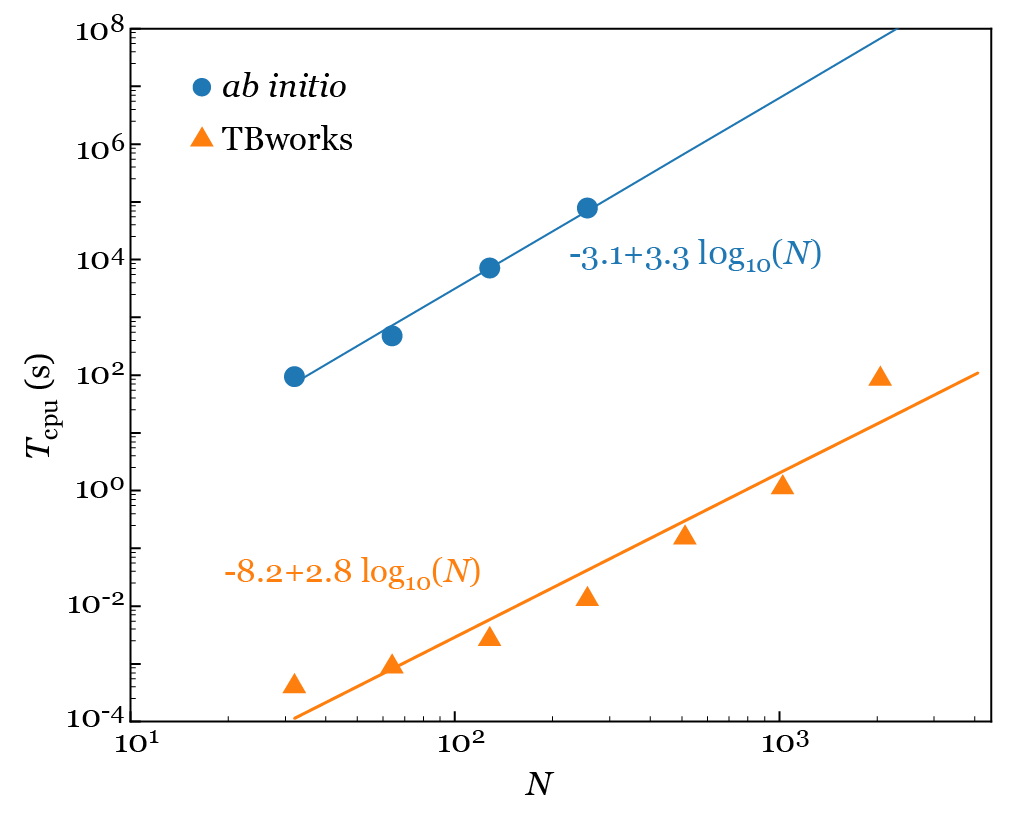}
	\protect\caption{(color online) Compute time $T_{\mathrm{cpu}}$ for calculating the electronic spectra of carbyne vs system size $N$ in \textit{ab initio} and TBworks calculations.  The \textit{ab initio} calculation is performed  at $\Gamma$ point \cite{Kresse1996}, with planewave basis cutoff at 400 eV and total energy convergence to within $10^{-4}$~eV per atom. The dots (\textit{ab initio}) and triangles (TBworks) mark the computation costs for different $N$. The solid lines are the least-square linear fittings between variable $\log_{10}(T_{\mathrm{cpu}})$ and $\log_{10}(N)$.}
	\label{fig:4}
\end{figure} 

With the accuracy perfectly preserved, we further investigate the efficiency of the TBworks model. The CPU times are obtained on a compute node equipped with 2 AMD EPYC-7452 CPUs (64 cores).  As shown in Fig.~\ref{fig:4}, due to the inevitable diagonalization procedure, the computation cost of both the TBworks and \textit{ab initio} calculations scale roughly as $\sim N^3$, where $N$ denotes the system size. However, the prefactor is different by more than 5 orders of magnitude. This is because in \textit{ab initio} calculations, diagonalization happens in solving the Kohn-Sham equations \cite{KohnSham1965}, where the one-electron wavefunctions are expanded over a large basis set of size proportional to but way larger than $N$. In contrast, the size of the TBworks Hamiltonian is the number of Wannier orbitals, which is generally of the same order of $N$.  As such, the dramatic speed-up in TBworks makes the accessible system size around 2 orders of magnitude larger than \textit{ab initio} calculations.

\begin{figure}[h]
	\centering
		\includegraphics[width=70 mm]{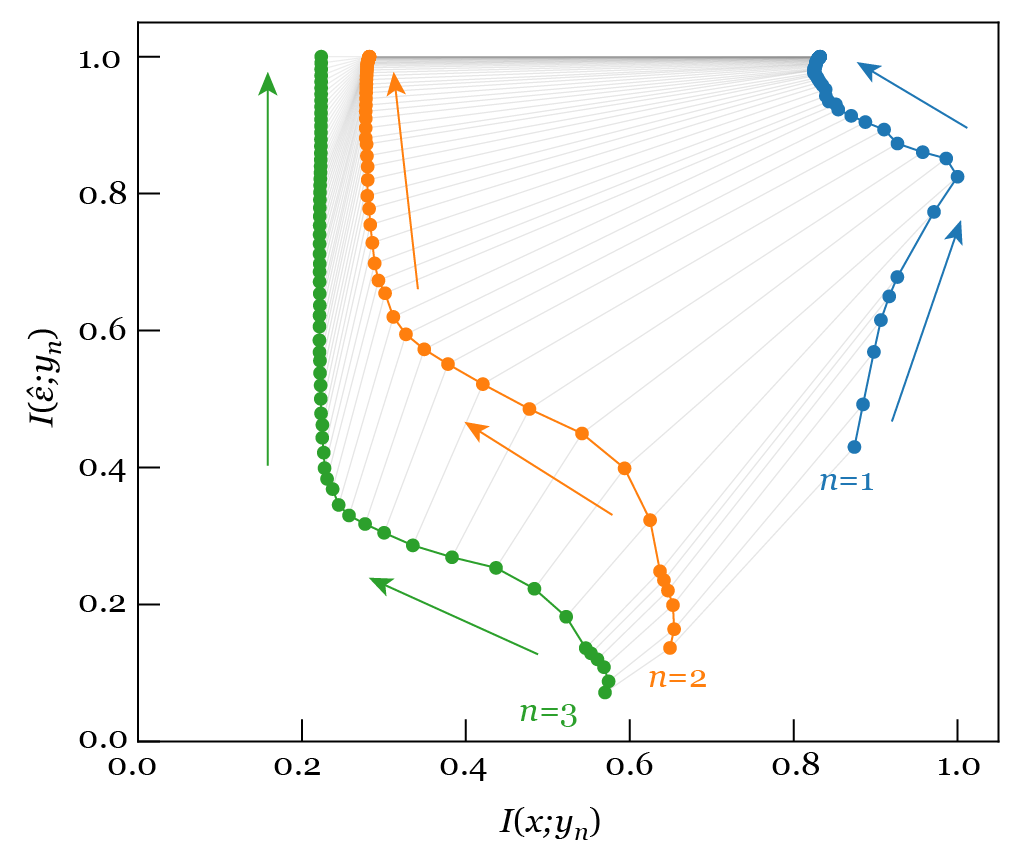}
	\protect\caption{(color online) Information plane loci of a 3-layer TBworks during training, showing the mutual information trajectories of a [200, 200, 200] neural network in 800 epoches. The colored solid arrows are guide-of-eye indications of the direction of progression for the three loci.}
	\label{fig:5}
\end{figure}

Further insights into the performance of the TBworks are gained by visualizing the training process on a reduced information plane \cite{Tishby_2015, Tishby2017}. The mutual information $I(a;b)$ is an asymmetric measure of the correlation between a pair of random variables $a$ and $b$, with distributions $P(a)$ and $P(b)$, respectively,
 \begin{equation}
 	I(a; b) = D_{\text{KL}}(P(a,b)||P(a)\otimes P(b))
 \end{equation}
where $D_{\text{KL}}$ is the Kullback-Leibler divergence and $P(a,b)$ is the joint distribution \cite{Kullback51}.
We compute the mutual information between the input $x$ (ionic coordinates inside a simulation box) and the output of the $n$-th layer $y_n$, and the mutual information between the training label $\hat \varepsilon$ and the output $y_n$. Thus we can visualize the evolution of the neural network on the reduced information plane $I(x,y_n)$-$I(\hat\varepsilon, y_n)$, as depicted in Fig.~\ref{fig:5}. We see that  $I(\hat \varepsilon,y_1)$ undergoes a continuous sharp increase throughout the training, which directly results from the optimization of the loss function, indicating continued improvement of model fitting. $I(x,y_1)$ increases slightly then decreases in the later stage. The decrease in $I(x,y_1)$ suggests a reduction in data redundancy, which is very important for the performance of the neural network. The dynamics of layers 2 and 3 are also bi-phasic, where redundancy elimination precedes the rapid model improvement.
The redundant information on $x$ arises from a large cut-off radius $R_{\mathrm{cut}}$ for the local chemical environment,  which includes atoms having little impact on the hopping amplitude. Therefore, the information plane analysis uncovers critical stages in the learning process of neural network during the training process, and offers leads for systematically optimizing the choice of setup parameter for TBworks.

\section{Applications to kinetic and equilibrium samplings}

The efficient, accurate and transferable access to $H(\br)$ offered by our TBworks provides a convenient computational tool for investigating the correlation between electronic and ionic degrees of freedom, which is particularly important for systems with strong electron-phonon coupling. 
As a first demonstration, TBworks is employed to evaluate correlation functions and corresponding physical properties of dynamical (both transient and equilibrium) systems. 
The detailed discussions centered around an application-oriented and task-specific example will help with an in-depth understanding of some of the issues in the implementation of the TBworks.
In the first example we compute the time-dependent spectral function of a kinetic process of soliton annihilation, whereupon  non-adiabatic effects are revealed. As a second  example, it is demonstrated  TBworks can be combined with path-integral molecular dynamics to compute the current-current correlation function and therefore the optical conductivity.
In this section, The evolution of ions is computed using ML force-field MD \cite{Linfeng_NIPS_2018} within the BO approximation. TBworks is used to predict the electronic Hamiltonian along ion trajectories from MD,  with which the  electrodynamics is obtained by numerically integrating the Schr\"odinger equation.  This approach allows us to capture non-BO effects on the electronic structure and responses of the electronic system.\footnote{Though no nonadiabatic effects on the ion dynamics is included in this approach.}

\begin{figure*}[htbp]
	\centering
		\includegraphics[width=130 mm]{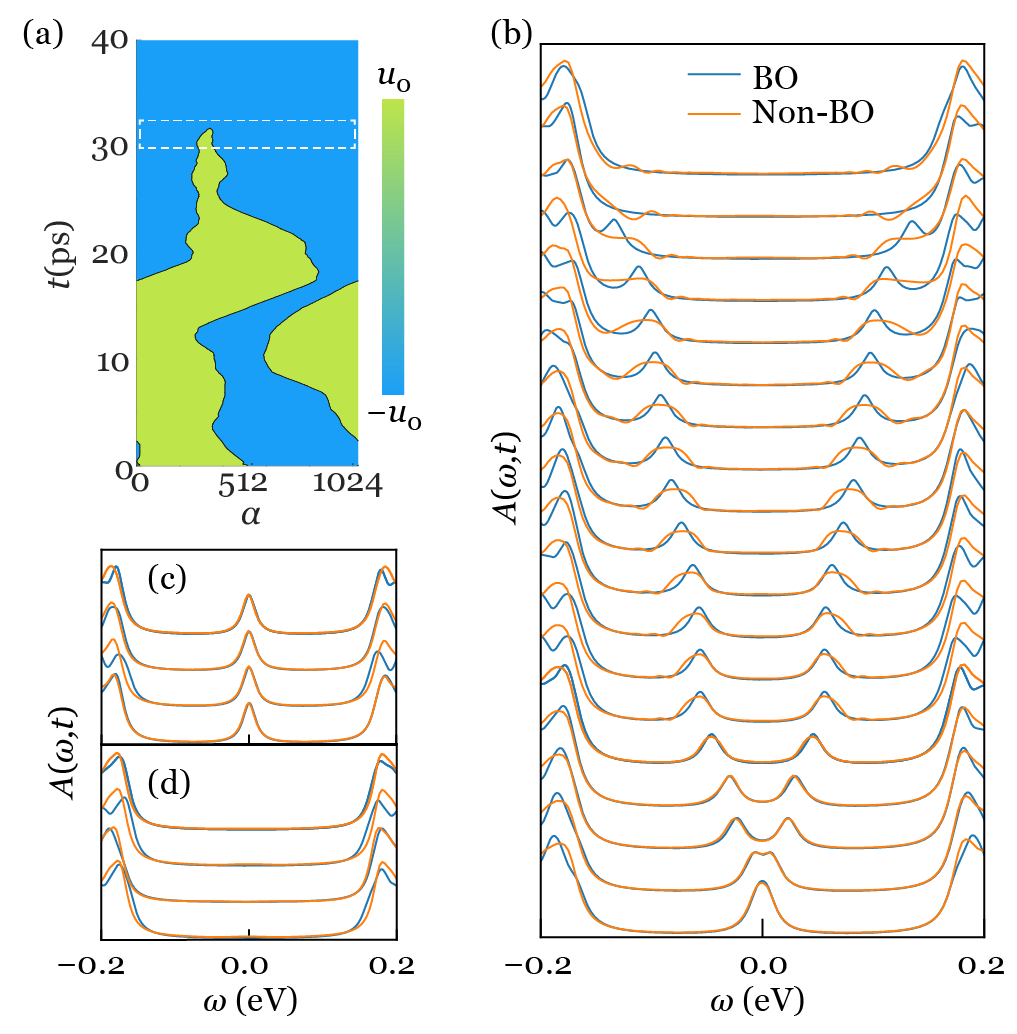}
	\protect\caption{(color online) Non-adiabatic dynamical effects on the electronic spectral functions during the soliton-antisoliton pair annihilation process. (a) The order parameters along a polyyne chain as a function of time in a molecular dynamics trajectory. The dashed rectangle indicated the soliton-antisoliton pair annihilation event. (b) Computed time-dependent electronic spectral functions for structures at every 20 fs in the annihilation event. (c)(d) Computed electronic spectral functions for structures before (c) and after (d) annihilation event. }
	\label{fig:6}
\end{figure*}

The soliton-antisoliton pair annihilation process is one such example, where the zero-energy electronic modes are gapped out dynamically. Non-adiabatic effects can be important but difficult to study in AIMD due to the high demand for computational resources. In order to study this kind of process, we interface our TBworks with the ML force field \cite{Linfeng_NIPS_2018} to perform highly efficient classical molecular dynamics (CMD) and path-integral molecular dynamics (PIMD) on carbyne by the large-scale atomic/molecular massively parallel simulator (LAMMPS) \cite{lammps1995}. The ML force field is generated by DeePMD-kit \cite{deepmdkit2018} using a $200\times 200\times 200$ deep neural network trained by 10000 snapshots of labeled AIMD data and tested by a validation set containing 1000 snapshots. The rms error is found to be within 0.01 eV/{\AA} for predicted forces, and $10^{-5}$ eV per atom for predicted energies, which means the accuracy of the ML force field is at the same level of the \textit{ab initio} calculation. Both the classical and path-integral MD simulations are performed in the canonical ensemble on a supercell  $N=1024$ with a time step of 0.2 femtoseconds (fs) at different temperatures. The Nose-Hoover thermostat \cite{Nose1984,Hoover1985,Tuckerman1993} is employed to control the temperatures in the canonical ensemble. In PIMD, the paths are discretized into 12 imaginary time slices.
Fig.~\ref{fig:6}(a) shows the order parameters $u_\alpha$ of the ionic configurations in the dynamical process where a soliton-antisoliton pair is annihilated in the classical MD at $T=1$~K.

To highlight the non-adiabatic effects in the soliton pair annihilation, we inspect the time-dependent electronic spectral functions. Because of the expedited access to TBworks Hamiltonian, we are able to let the electronic system evolve according to 
$\mathcal{T}e^{-\ii \int_{t_0}^{t}H\left(r(t^\prime)\right)dt^\prime}$
 ($\hbar=1$). Doing so explicitly takes ion dynamics into account and is manifestly beyond the BO approximation. As ion dynamics breaks time homogeneity for electrons, we define a time-dependent spectral function as
\begin{eqnarray}
A(\omega, t)= \frac{\ii}{2\pi N} \hat F  \operatorname{Tr}\left[G^R (t+\tau,t)- G^A(t+\tau,t)\right]
\label{eq.spectral_t}
\end{eqnarray}
where $G^{A/R}_{\alpha\beta}(t+\tau,t)=\pm\ii\Theta(\mp \tau)\langle\{c_{\alpha}(t+\tau), c_{\beta}^{\dagger}(t)\}\rangle$ are advanced/retarded Green's functions, and $\hat F$ means a Fourier transform w.r.t. the time difference $\tau$ \cite{Nghiem17}. At different $t$, the advanced and retarded Green's functions are evaluated by computing the expectation of the commutator $\{c_{\alpha}(t+\tau), c_{\beta}^{\dagger}(t)\}$ for $\tau >0$  and $\tau <0$, respectively, with  $|\tau|\le 600$ for an energy resolution  of $\sim 3$ meV  ($\sim 1\%$ of the CDW gap). As the exact propagator from the TBworks Hamiltonian can be quickly computed, the  $A(\omega, t)$ so obtained can be non-adiabatically calculated along a trajectory in the MD simulation. As a comparison, the static electronic spectral function within the BO approximation for configuration at $t$ is calculated using instantaneous energy spectra, with 
$
A_{\text{BO}} (\omega,t) = \frac{1}{N_{k}}\sum_{nk} \delta(\omega - \varepsilon_{nk}(t))
$. 

We focus on the soliton-antisoliton pair annihilation event indicated by the dashed rectangle in Fig.~\ref{fig:6}(a). During this event, two kinks meet and annihilate, and the polyyne-B phase completely disappears. At the same time, the degenerate zero-energy modes localized at the kinks become hybridized and a gap gradually develops. The question is whether the BO approximation is accurate enough to describe the electronic states for the dynamical annihilation process. As a comparison, we first examined the electronic spectral functions for structures well before (Fig.~\ref{fig:6}(c)) and after (Fig.~\ref{fig:6}(d)) the annihilation event. Prior to the annihilation event where the soliton and anti-soliton are well separated, both dynamical (beyond the BO approximation) and static (within the BO approximation) spectral functions give the same topological protected zero-energy peaks, as shown in Fig.~\ref{fig:6}(c), where little difference between the dynamical and static spectral functions is discernible. Post annihilation, the carbyne chain adopts a homogeneous polyyne-A structure, where both dynamical and static spectral functions present the CDW gap, as shown in Fig.~\ref{fig:6}(d). Again, little difference between the dynamical and static spectral functions is uncovered, because the system is in a (transient) stationary state.

 Interestingly, in the course of this annihilation event, the dynamical and static time-dependent spectral functions for the annihilation process display remarkable departures, as shown in Fig.~\ref{fig:6}(b). Both dynamical and static spectral functions show the splitting of the zero-energy modes into two peaks, which subsequently merge into the conduction and valence bands, respectively, as the soliton pair approaching annihilation. However,  the low-energy peaks in the dynamical spectral functions show significantly larger broadening than those in the static spectral functions. Thus, although the dynamical effects shows little renormalization on the spectral functions before and after the annihilation event, the non-adiabatic dynamical effects are clearly visible in the splitting of the topological in-gap modes in our calculations.

The nonadiabatic effects on the low-energy modes can be revealed in the spectral function projected on atomic sites,
 \begin{eqnarray}
	A(\omega, \alpha)= \frac{\ii}{2\pi N}\left[G^R_{\alpha \alpha} (\omega)- G^A_{\alpha \alpha}(\omega)\right]
\end{eqnarray}
where $G_{\alpha \alpha}^{A/R}(\omega)$ are the on-site advanced/retarded Green’s functions and $\alpha$ is the atomic index. (For brevity, we leave out the index $t$ shown in Eq.\ref{eq.spectral_t}.) For comparison, the static spectral function in the BO approximation is calculated from the instantaneous energy spectra and wavefunctions with \\
$
A_{\text{BO}}(\omega,\alpha)= \frac{1}{N_{k}}\sum_{nk} \delta(\omega - \varepsilon_{nk}) \left\lvert a_{n\alpha}\right\rvert^2$.
where the $a_{n\alpha}$ is the expansion coefficient of $n$-th wavefunction in the atomic orbital basis. 

\begin{figure}[h]
	\centering
		\includegraphics[width=80 mm]{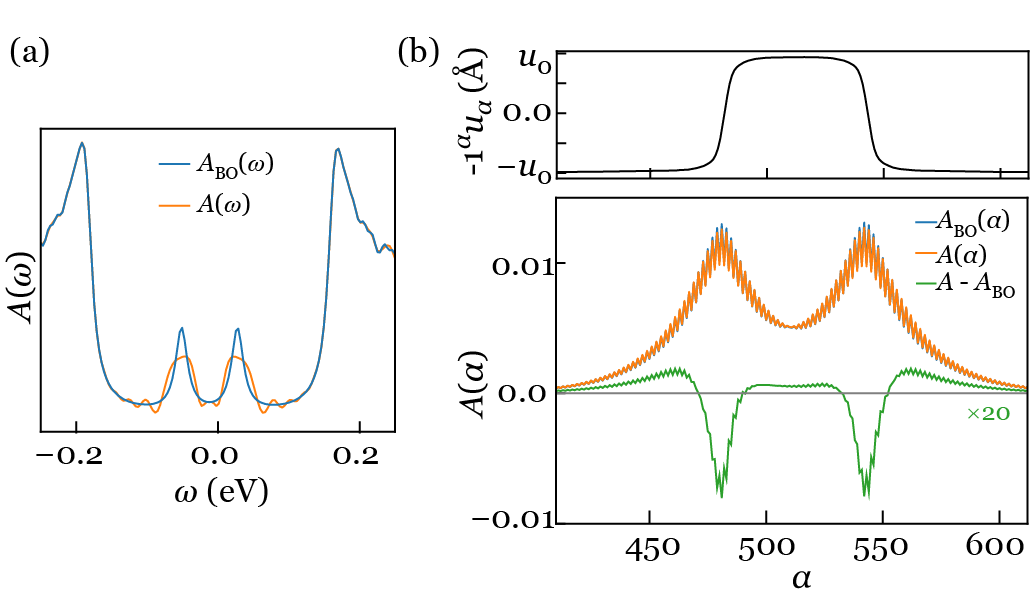}
	\protect\caption{(color online) Spectral functions of the low-energy electronic modes. (a) Calculated static and dynamical spectral functions. (b) Upper: the order parameter of a carbyne chain with a soliton-antisoliton pair. Lower: static (BO) and dynamical spectral weights along the carbyne chain. Also shown (green line) is their difference. }
	\label{fig:7}
\end{figure}

Both dynamical and static spectral functions are calculated for a carbyne configuration where the soliton and anti-soliton are moving toward each other before annihilation simulated in MD with $N=1024$ and $T=1$~K. As order parameter plotted in upper panel of Fig.~\ref{fig:7}(b) indicates, the soliton and anti-soliton are about 60 sites apart and there is a significant coupling. As shown in Fig.~\ref{fig:7}(a) the low-energy modes are split because of the coupling, and the modes in the dynamical spectral function are clearly broadened compared to the static ones due to nonadiabatic effects.

The origin of the broadening of the low-energy peak is the retardation of electronic responses as ions move, which is by definition absent in the BO calculation. This can be demonstrated by inspecting the site-projected spectral weights. The contribution from the atomic orbitals to the low-energy modes is measured by 
$A(\alpha)=\int_{-0.1}^{0.1} d\omega A(\omega,\alpha)$. The lower panel of Fig.~\ref{fig:7}(b) shows that both the static and dynamical spectral functions peaked around the soliton and anti-soliton, with significant overlap commensurate with the finite coupling. By taking the difference between the two spectral functions, we see that the dynamical spectral function is more delocalized, carrying especially a larger weight on the tails of the inward moving solitons. This in fact reflects the retardation effects that electrons do not immediately follow the moving solitons, which cannot be captured by the BO calculations. We emphasize that it is very difficult, if possible at all, to produce these results directly from \textit{ab initio} calculations for two reasons. First, the pair annihilation process requires very long carbon chain in the simulation. Second, the non-adiabatic effects require a time-dependent electronic evolution concomitant to the MD, which is an added challenge on top of the already taxing requirement of large model system. But with TBworks, this type of calculations are performed with much reduced computational costs.

Now we move on to examine some of the equilibrium electronic correlation functions based on the TBworks, again with carbyne as the model material, using classical and path integral MD. In order to compute experimentally accessible electronic properties, retarded correlation functions of the form \cite{mahan2013}
\begin{equation}
	G_{AB}^{R}(t-t')=-\ii \lambda\,  \Theta(t)\operatorname{Tr}\left\{\varrho_{\text{ion}}(\vec r') \langle[A(t), B(t')]_s\rangle\right\}
\end{equation}
usually need to be computed, where $s=\pm$ for Fermion/Boson and $\lambda$ is a numerical constant.
In an equilibrium ensemble, the sampling according to ionic density matrix $\varrho_{\text{ion}}(\vec r(t'))$ is automatically offered by the trajectory from molecular dynamics. Anticipating little non-adiabatic effects, we use the approximate propagators: $U(t-t')\approx e^{-\ii H(t')(t-t')}$. Then the electronic spectral function is computed with $A\rightarrow \hat \psi, B\rightarrow \hat\psi ^\dagger$ and $\lambda =\ii/2\pi N$. To compute the electronic conductivity, $A, B\rightarrow j= -env$ the current operator. Then the dissipative part of the conductivity is given by 
\begin{equation}
	\operatorname{Re}\sigma(\omega) = \operatorname{Re}\left[\frac{\ii}{\omega}
	 G_{jj}^R(\omega)\right].
\end{equation}
For the Fourier transformation, time values $ t-t' \in [-600, 600]$ fs are used with 0.1 fs increment. Further details of the calculation of these correlation functions can be found in Appendix section B.

\begin{figure}[h]
	\centering
		\includegraphics[width=80 mm]{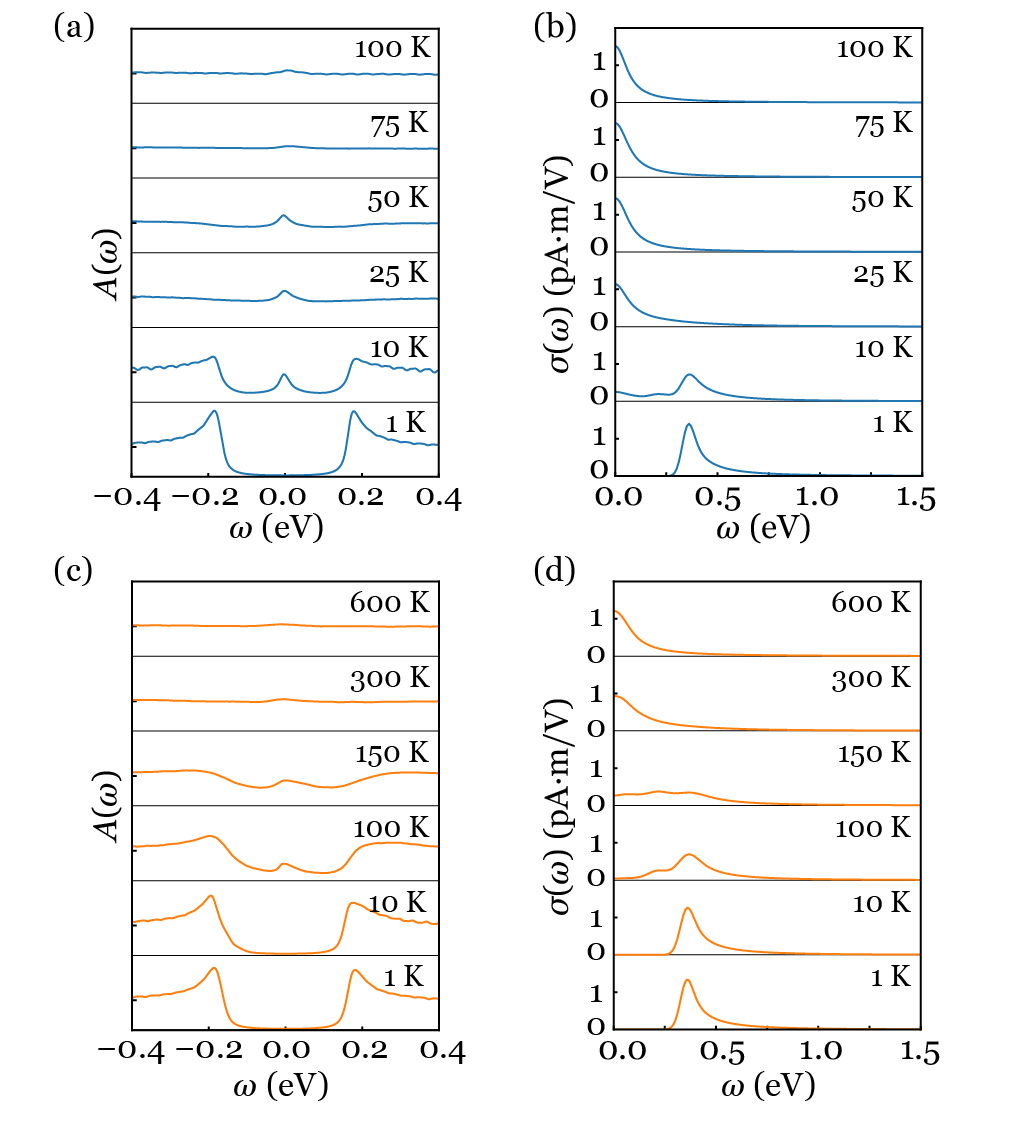}
	\protect\caption{(color online) The averaged electronic spectral function $A(\omega)$ and optical conductivity $\sigma(\omega)$ for supercell $N=1024$ at different temperatures.  Spectral function (a)(c) and optical conductivity (b)(d) averaged in the canonical ensemble sampled by PIMD (blue lines) and CMD (orange lines).}
	\label{fig:8}
\end{figure} 

The computed spectral function and optical conductivity for carbyne at finite temperatures are shown in Fig. \ref{fig:8}. We see that at $T=1$~K, the spectral functions from both path-integral and classical simulations show a CDW gap $\Delta \sim$ $0.27$~eV, indicating the equilibrium configuration of carbyne is in the polyyne structure. The optical conductivity also shows an onset at $\omega = \Delta$ at this temperature. At $T=10$~K, the spectral function remains gapped in CMD, indicating absence of soliton. In contrast, PIMD shows a zero-energy peak in spectral function, which suggests quantum effects promotes the generation of soliton-antisoliton pairs. As shown in  Fig.~A-1, the energy barrier for generation of a soliton-antisoliton pair from the polyyne structure is $\sim 80$~meV. Therefore, at low temperature $10$~K, the soliton-antisoliton pair is hardly sampled in CMD. However, in PIMD due to quantum dynamical effects, the creation of soliton-antisoliton pair can be observed. 

When the temperature is raised to moderately high value $100$~K, in PIMD, the CDW gap and the zero-energy mode disappear and optical conductivity shows the single peak at $\omega=0$, which is commensurate with the polycumulene phase. At the same temperature in CMD, the zero-energy peak in spectral function indicates presence of solitons. Above $300$~K, in the classical molecular dynamics, the CDW melts and polycumulene is the equilibrium structure. Both spectral function and optical conductivity present the sequence of phase transitions in carbyne from polyyne to soliton and finally to polycumulene in CMD and PIMD. Due to the quantum dynamical effects, the soliton and polycumulene phase transitions happen at lower temperatures in PIMD than that in CMD, which could be detected by measuring the spectral function and/or optical conductivity experimentally. All these results exhibit the efficiency of our TBworks in calculating correlation functions that can be routinely measured in experiment to monitor the dynamics and phase transition but require large-scale and long-time simulations.

\section{Summary}
In this work, we have developed a neural network-based predictive representation method, which predicts tight-binding Hamiltonians for modeling electronic structures. This method provides an efficient approach to configuration-dependent Hamiltonians by directly supervised observation of the \textit{ab initio} energy spectra. This enables an expedited access to electronic structures at dramatically reduced computational costs during the atomic configurations sampling, in subsequent MD or Monte Carlo. Considering the example of carbyne, we demonstrate the accuracy of our method in predicting the electronic spectra is comparable to that of the \textit{ab initio} calculations. The TBworks also shows a high degree of transferability, accurately predicting spectra for configurations drawn from unseen temperatures and system sizes. 

By interfacing our TBworks with a ML force field it is demonstrated that we can efficiently sample the electronic structure of carbyne in MD to compute correlation functions. This enables the computation of Green's functions and the time-dependent spectral functions in the dynamical annihilation of a soliton-antisoliton pair. When comparing our results with those from the adiabatic approximation, it is found that the low-energy edge modes are significant renormalized due to the non-adiabatic dynamical effects. We have also applied the TBworks sampling of electronic structure in a PIMD, where the optical conductivity is computed by sampling the current-current correlation function.

In summary, the TBworks offers an expedited access to electronic structure of a dynamical system with \textit{ab initio} accuracy, based on an accurate and transferable representation of electronic structures from AIMD. This entails a dramatic speedup and reduction of computational cost, when combined with ML force field MD, in computing electronic correlation functions and associated physical properties  both at equilibrium and in a transient process. This will be especially useful for problems where large-scale molecular dynamics and long sampling times are essential, as we have demonstrated in the non-adiabatic process of soliton annihilation. Evidently, this approach will significantly enhance our capability and broaden the scope in the modeling of electronic structure and processes in materials at finite temperatures or in kinetic phenomena. 

Furthermore, we believe that the TBworks approach can have far-reaching potential applications in materials simulations. Because of the accurate and efficient access to electronic Hamiltonian as a function of ion configuration it offers, the approach will be especially useful to explore the properties emerging from dynamical processes and from electron-phonon coupling (e.g., phonon-assisted indirect optical absorption). Moreover, the high efficiency of TBworks in comparison with full \textit{ab initio} electronic structure calculation will enable large-scale and long-time simulations, which can offer a direct comparison to experimentally relevant temporal and spatial scales. Clearly, further developments to deal with more complex systems are worthy goals to pursue in the future. Extension of the TBworks approach to higher dimensional materials is currently underway. It is also important to incorporate the full symmetry of the system and implement symmetry-adapted TBworks \cite{Linfeng_NIPS_2018}, which will further improve the training efficiency and model transferability.

\section*{Conflict of interest}
The authors declare that they have no conflict of interest.

\section*{Acknowledgments}
  This work is supported by the National Natural Science Foundation of China (11725415 and 11934001), the Ministry of Science and Technology of  China (2018YFA0305601 and 2016YFA0301004), and by the Strategic Priority Research Program of the Chinese Academy of Sciences  (XDB28000000). One of the authors (Linfeng Zhang) has been supported in part by the Center for Chemistry in Solution and at Interfaces (CSI) at Princeton University, funded by the DOE Award DE-SC0019394.

\section*{Author contributions}
All authors made indispensable contributions to the research, and wrote and critically revised the manuscript together. 



\begin{thebibliography}{38}%
	\makeatletter
	\providecommand \@ifxundefined [1]{%
		\@ifx{#1\undefined}
	}%
	\providecommand \@ifnum [1]{%
		\ifnum #1\expandafter \@firstoftwo
		\else \expandafter \@secondoftwo
		\fi
	}%
	\providecommand \@ifx [1]{%
		\ifx #1\expandafter \@firstoftwo
		\else \expandafter \@secondoftwo
		\fi
	}%
	\providecommand \natexlab [1]{#1}%
	\providecommand \enquote  [1]{``#1''}%
	\providecommand \bibnamefont  [1]{#1}%
	\providecommand \bibfnamefont [1]{#1}%
	\providecommand \citenamefont [1]{#1}%
	\providecommand \href@noop [0]{\@secondoftwo}%
	\providecommand \href [0]{\begingroup \@sanitize@url \@href}%
	\providecommand \@href[1]{\@@startlink{#1}\@@href}%
	\providecommand \@@href[1]{\endgroup#1\@@endlink}%
	\providecommand \@sanitize@url [0]{\catcode `\\12\catcode `\$12\catcode
		`\&12\catcode `\#12\catcode `\^12\catcode `\_12\catcode `\%12\relax}%
	\providecommand \@@startlink[1]{}%
	\providecommand \@@endlink[0]{}%
	\providecommand \url  [0]{\begingroup\@sanitize@url \@url }%
	\providecommand \@url [1]{\endgroup\@href {#1}{\urlprefix }}%
	\providecommand \urlprefix  [0]{URL }%
	\providecommand \Eprint [0]{\href }%
	\providecommand \doibase [0]{https://doi.org/}%
	\providecommand \selectlanguage [0]{\@gobble}%
	\providecommand \bibinfo  [0]{\@secondoftwo}%
	\providecommand \bibfield  [0]{\@secondoftwo}%
	\providecommand \translation [1]{[#1]}%
	\providecommand \BibitemOpen [0]{}%
	\providecommand \bibitemStop [0]{}%
	\providecommand \bibitemNoStop [0]{.\EOS\space}%
	\providecommand \EOS [0]{\spacefactor3000\relax}%
	\providecommand \BibitemShut  [1]{\csname bibitem#1\endcsname}%
	\let\auto@bib@innerbib\@empty
	\bibitem [{\citenamefont {Car}\ and\ \citenamefont
		{Parrinello}(1985)}]{Parrinello1985}%
	\BibitemOpen
	\bibfield  {author} {\bibinfo {author} {\bibfnamefont {R.}~\bibnamefont
			{Car}}\ and\ \bibinfo {author} {\bibfnamefont {M.}~\bibnamefont
			{Parrinello}},\ }\bibfield  {title} {\bibinfo {title} {Unified approach for
			molecular dynamics and density-functional theory},\ }\href@noop {} {\bibfield
		{journal} {\bibinfo  {journal} {Phys. Rev. Lett.}\ }\textbf {\bibinfo
			{volume} {55}},\ \bibinfo {pages} {2471} (\bibinfo {year}
		{1985})}\BibitemShut {NoStop}%
	\bibitem [{\citenamefont {Marx}\ and\ \citenamefont
		{Hutter}(2009)}]{Hutter2009}%
	\BibitemOpen
	\bibfield  {author} {\bibinfo {author} {\bibfnamefont {D.}~\bibnamefont
			{Marx}}\ and\ \bibinfo {author} {\bibfnamefont {J.}~\bibnamefont {Hutter}},\
	}\href@noop {} {\emph {\bibinfo {title} {Ab Initio Molecular Dynamics: Basic
				Theory and Advanced Methods}}}\ (\bibinfo  {publisher} {Cambridge University
		Press},\ \bibinfo {year} {2009})\BibitemShut {NoStop}%
	\bibitem [{\citenamefont {LeCun}\ \emph {et~al.}(2015)\citenamefont {LeCun},
		\citenamefont {Bengio},\ and\ \citenamefont {Hinton}}]{LeCun2015}%
	\BibitemOpen
	\bibfield  {author} {\bibinfo {author} {\bibfnamefont {Y.}~\bibnamefont
			{LeCun}}, \bibinfo {author} {\bibfnamefont {Y.}~\bibnamefont {Bengio}},\ and\
		\bibinfo {author} {\bibfnamefont {G.}~\bibnamefont {Hinton}},\ }\bibfield
	{title} {\bibinfo {title} {Deep learning},\ }\href@noop {} {\bibfield
		{journal} {\bibinfo  {journal} {Nature}\ }\textbf {\bibinfo {volume} {521}},\
		\bibinfo {pages} {436} (\bibinfo {year} {2015})}\BibitemShut {NoStop}%
	\bibitem [{\citenamefont {Behler}\ and\ \citenamefont
		{Parrinello}(2007)}]{Behler2007}%
	\BibitemOpen
	\bibfield  {author} {\bibinfo {author} {\bibfnamefont {J.}~\bibnamefont
			{Behler}}\ and\ \bibinfo {author} {\bibfnamefont {M.}~\bibnamefont
			{Parrinello}},\ }\bibfield  {title} {\bibinfo {title} {Generalized
			neural-network representation of high-dimensional potential-energy
			surfaces},\ }\href@noop {} {\bibfield  {journal} {\bibinfo  {journal} {Phys.
				Rev. Lett.}\ }\textbf {\bibinfo {volume} {98}},\ \bibinfo {pages} {146401}
		(\bibinfo {year} {2007})}\BibitemShut {NoStop}%
	\bibitem [{\citenamefont {Sch{\"u}tt}\ \emph {et~al.}(2017)\citenamefont
		{Sch{\"u}tt}, \citenamefont {Arbabzadah}, \citenamefont {Chmiela},
		\citenamefont {Müller},\ and\ \citenamefont {Tkatchenko}}]{Schutt2017}%
	\BibitemOpen
	\bibfield  {author} {\bibinfo {author} {\bibfnamefont {K.~T.}\ \bibnamefont
			{Sch{\"u}tt}}, \bibinfo {author} {\bibfnamefont {F.}~\bibnamefont
			{Arbabzadah}}, \bibinfo {author} {\bibfnamefont {S.}~\bibnamefont {Chmiela}},
		\bibinfo {author} {\bibfnamefont {K.~R.}\ \bibnamefont {Müller}},\ and\
		\bibinfo {author} {\bibfnamefont {A.}~\bibnamefont {Tkatchenko}},\ }\bibfield
	{title} {\bibinfo {title} {Quantum-chemical insights from deep tensor neural
			networks},\ }\href@noop {} {\bibfield  {journal} {\bibinfo  {journal} {Nat.
				Commun.}\ }\textbf {\bibinfo {volume} {8}},\ \bibinfo {pages} {13890}
		(\bibinfo {year} {2017})}\BibitemShut {NoStop}%
	\bibitem [{\citenamefont {Schütt}\ \emph {et~al.}(2018)\citenamefont
		{Schütt}, \citenamefont {Sauceda}, \citenamefont {Kindermans}, \citenamefont
		{Tkatchenko},\ and\ \citenamefont {Müller}}]{Schutt2018}%
	\BibitemOpen
	\bibfield  {author} {\bibinfo {author} {\bibfnamefont {K.~T.}\ \bibnamefont
			{Schütt}}, \bibinfo {author} {\bibfnamefont {H.~E.}\ \bibnamefont
			{Sauceda}}, \bibinfo {author} {\bibfnamefont {P.-J.}\ \bibnamefont
			{Kindermans}}, \bibinfo {author} {\bibfnamefont {A.}~\bibnamefont
			{Tkatchenko}},\ and\ \bibinfo {author} {\bibfnamefont {K.-R.}\ \bibnamefont
			{Müller}},\ }\bibfield  {title} {\bibinfo {title} {Schnet – a deep
			learning architecture for molecules and materials},\ }\href@noop {}
	{\bibfield  {journal} {\bibinfo  {journal} {J. Chem. Phys.}\ }\textbf
		{\bibinfo {volume} {148}},\ \bibinfo {pages} {241722} (\bibinfo {year}
		{2018})}\BibitemShut {NoStop}%
	\bibitem [{\citenamefont {Zhang}\ \emph
		{et~al.}(2018{\natexlab{a}})\citenamefont {Zhang}, \citenamefont {Han},
		\citenamefont {Wang}, \citenamefont {Car},\ and\ \citenamefont
		{E}}]{Linfeng2018}%
	\BibitemOpen
	\bibfield  {author} {\bibinfo {author} {\bibfnamefont {L.}~\bibnamefont
			{Zhang}}, \bibinfo {author} {\bibfnamefont {J.}~\bibnamefont {Han}}, \bibinfo
		{author} {\bibfnamefont {H.}~\bibnamefont {Wang}}, \bibinfo {author}
		{\bibfnamefont {R.}~\bibnamefont {Car}},\ and\ \bibinfo {author}
		{\bibfnamefont {W.}~\bibnamefont {E}},\ }\bibfield  {title} {\bibinfo {title}
		{Deep potential molecular dynamics: A scalable model with the accuracy of
			quantum mechanics},\ }\href@noop {} {\bibfield  {journal} {\bibinfo
			{journal} {Phys. Rev. Lett.}\ }\textbf {\bibinfo {volume} {120}},\ \bibinfo
		{pages} {143001} (\bibinfo {year} {2018}{\natexlab{a}})}\BibitemShut
	{NoStop}%
	\bibitem [{\citenamefont {Zhang}\ \emph
		{et~al.}(2018{\natexlab{b}})\citenamefont {Zhang}, \citenamefont {Han},
		\citenamefont {Wang}, \citenamefont {Saidi}, \citenamefont {Car},\ and\
		\citenamefont {E}}]{Linfeng_NIPS_2018}%
	\BibitemOpen
	\bibfield  {author} {\bibinfo {author} {\bibfnamefont {L.}~\bibnamefont
			{Zhang}}, \bibinfo {author} {\bibfnamefont {J.}~\bibnamefont {Han}}, \bibinfo
		{author} {\bibfnamefont {H.}~\bibnamefont {Wang}}, \bibinfo {author}
		{\bibfnamefont {W.}~\bibnamefont {Saidi}}, \bibinfo {author} {\bibfnamefont
			{R.}~\bibnamefont {Car}},\ and\ \bibinfo {author} {\bibfnamefont
			{W.}~\bibnamefont {E}},\ }\bibfield  {title} {\bibinfo {title} {End-to-end
			symmetry preserving inter-atomic potential energy model for finite and
			extended systems},\ }\href@noop {} {\bibfield  {journal} {\bibinfo  {journal}
			{Adv. Neural Inf. Process. Syst.}\ }\textbf {\bibinfo {volume} {32}},\
		\bibinfo {pages} {4436} (\bibinfo {year} {2018}{\natexlab{b}})}\BibitemShut
	{NoStop}%
	\bibitem [{\citenamefont {Ben~Mahmoud}\ \emph {et~al.}(2020)\citenamefont
		{Ben~Mahmoud}, \citenamefont {Anelli}, \citenamefont {Cs\'anyi},\ and\
		\citenamefont {Ceriotti}}]{Mahmoud2020}%
	\BibitemOpen
	\bibfield  {author} {\bibinfo {author} {\bibfnamefont {C.}~\bibnamefont
			{Ben~Mahmoud}}, \bibinfo {author} {\bibfnamefont {A.}~\bibnamefont {Anelli}},
		\bibinfo {author} {\bibfnamefont {G.}~\bibnamefont {Cs\'anyi}},\ and\
		\bibinfo {author} {\bibfnamefont {M.}~\bibnamefont {Ceriotti}},\ }\bibfield
	{title} {\bibinfo {title} {Learning the electronic density of states in
			condensed matter},\ }\href@noop {} {\bibfield  {journal} {\bibinfo  {journal}
			{Phys. Rev. B}\ }\textbf {\bibinfo {volume} {102}},\ \bibinfo {pages}
		{235130} (\bibinfo {year} {2020})}\BibitemShut {NoStop}%
	\bibitem [{\citenamefont {Chandrasekaran}\ \emph {et~al.}(2019)\citenamefont
		{Chandrasekaran}, \citenamefont {Kamal}, \citenamefont {Batra}, \citenamefont
		{Kim}, \citenamefont {Chen},\ and\ \citenamefont
		{Ramprasad}}]{Chandrasekaran2019}%
	\BibitemOpen
	\bibfield  {author} {\bibinfo {author} {\bibfnamefont {A.}~\bibnamefont
			{Chandrasekaran}}, \bibinfo {author} {\bibfnamefont {D.}~\bibnamefont
			{Kamal}}, \bibinfo {author} {\bibfnamefont {R.}~\bibnamefont {Batra}},
		\bibinfo {author} {\bibfnamefont {C.}~\bibnamefont {Kim}}, \bibinfo {author}
		{\bibfnamefont {L.}~\bibnamefont {Chen}},\ and\ \bibinfo {author}
		{\bibfnamefont {R.}~\bibnamefont {Ramprasad}},\ }\bibfield  {title} {\bibinfo
		{title} {Solving the electronic structure problem with machine learning},\
	}\href@noop {} {\bibfield  {journal} {\bibinfo  {journal} {Npj Comput.
				Mater.}\ }\textbf {\bibinfo {volume} {5}},\ \bibinfo {pages} {22} (\bibinfo
		{year} {2019})}\BibitemShut {NoStop}%
	\bibitem [{\citenamefont {Schütt}\ \emph {et~al.}(2019)\citenamefont
		{Schütt}, \citenamefont {Gastegger}, \citenamefont {Tkatchenko},
		\citenamefont {Müller},\ and\ \citenamefont {Maurer}}]{Schuett2019}%
	\BibitemOpen
	\bibfield  {author} {\bibinfo {author} {\bibfnamefont {K.~T.}\ \bibnamefont
			{Schütt}}, \bibinfo {author} {\bibfnamefont {M.}~\bibnamefont {Gastegger}},
		\bibinfo {author} {\bibfnamefont {A.}~\bibnamefont {Tkatchenko}}, \bibinfo
		{author} {\bibfnamefont {K.-R.}\ \bibnamefont {Müller}},\ and\ \bibinfo
		{author} {\bibfnamefont {R.~J.}\ \bibnamefont {Maurer}},\ }\bibfield  {title}
	{\bibinfo {title} {Unifying machine learning and quantum chemistry with a
			deep neural network for molecular wavefunctions},\ }\href@noop {} {\bibfield
		{journal} {\bibinfo  {journal} {Nat. Commun.}\ }\textbf {\bibinfo {volume}
			{10}},\ \bibinfo {pages} {5024} (\bibinfo {year} {2019})}\BibitemShut
	{NoStop}%
	\bibitem [{\citenamefont {Brouder}\ \emph {et~al.}(2007)\citenamefont
		{Brouder}, \citenamefont {Panati}, \citenamefont {Calandra}, \citenamefont
		{Mourougane},\ and\ \citenamefont {Marzari}}]{Brouder07}%
	\BibitemOpen
	\bibfield  {author} {\bibinfo {author} {\bibfnamefont {C.}~\bibnamefont
			{Brouder}}, \bibinfo {author} {\bibfnamefont {G.}~\bibnamefont {Panati}},
		\bibinfo {author} {\bibfnamefont {M.}~\bibnamefont {Calandra}}, \bibinfo
		{author} {\bibfnamefont {C.}~\bibnamefont {Mourougane}},\ and\ \bibinfo
		{author} {\bibfnamefont {N.}~\bibnamefont {Marzari}},\ }\bibfield  {title}
	{\bibinfo {title} {Exponential localization of {Wannier} functions in
			insulators},\ }\href@noop {} {\bibfield  {journal} {\bibinfo  {journal}
			{Phys. Rev. Lett.}\ }\textbf {\bibinfo {volume} {98}},\ \bibinfo {pages}
		{046402} (\bibinfo {year} {2007})}\BibitemShut {NoStop}%
	\bibitem [{\citenamefont {Cloizeaux}(1964)}]{Cloizeaux1964}%
	\BibitemOpen
	\bibfield  {author} {\bibinfo {author} {\bibfnamefont {J.~D.}\ \bibnamefont
			{Cloizeaux}},\ }\bibfield  {title} {\bibinfo {title} {Analytical properties
			of $n$-dimensional energy bands and {Wannier} functions},\ }\href@noop {}
	{\bibfield  {journal} {\bibinfo  {journal} {Phys. Rev.}\ }\textbf {\bibinfo
			{volume} {135}},\ \bibinfo {pages} {A698} (\bibinfo {year}
		{1964})}\BibitemShut {NoStop}%
	\bibitem [{\citenamefont {Nenciu}(1983)}]{nenciu1983}%
	\BibitemOpen
	\bibfield  {author} {\bibinfo {author} {\bibfnamefont {G.}~\bibnamefont
			{Nenciu}},\ }\bibfield  {title} {\bibinfo {title} {Existence of the
			exponentially localised {Wannier} functions},\ }\href@noop {} {\bibfield
		{journal} {\bibinfo  {journal} {Comm. Math. Phys.}\ }\textbf {\bibinfo
			{volume} {91}},\ \bibinfo {pages} {81} (\bibinfo {year} {1983})}\BibitemShut
	{NoStop}%
	\bibitem [{\citenamefont {Marzari}\ \emph {et~al.}(2012)\citenamefont
		{Marzari}, \citenamefont {Mostofi}, \citenamefont {Yates}, \citenamefont
		{Souza},\ and\ \citenamefont {Vanderbilt}}]{Marzari12}%
	\BibitemOpen
	\bibfield  {author} {\bibinfo {author} {\bibfnamefont {N.}~\bibnamefont
			{Marzari}}, \bibinfo {author} {\bibfnamefont {A.~A.}\ \bibnamefont
			{Mostofi}}, \bibinfo {author} {\bibfnamefont {J.~R.}\ \bibnamefont {Yates}},
		\bibinfo {author} {\bibfnamefont {I.}~\bibnamefont {Souza}},\ and\ \bibinfo
		{author} {\bibfnamefont {D.}~\bibnamefont {Vanderbilt}},\ }\bibfield  {title}
	{\bibinfo {title} {Maximally localized {Wannier} functions: Theory and
			applications},\ }\href@noop {} {\bibfield  {journal} {\bibinfo  {journal}
			{Rev. Mod. Phys.}\ }\textbf {\bibinfo {volume} {84}},\ \bibinfo {pages}
		{1419} (\bibinfo {year} {2012})}\BibitemShut {NoStop}%
	\bibitem [{\citenamefont {Kingma}\ and\ \citenamefont {Ba}(2015)}]{Adam2014}%
	\BibitemOpen
	\bibfield  {author} {\bibinfo {author} {\bibfnamefont {D.~P.}\ \bibnamefont
			{Kingma}}\ and\ \bibinfo {author} {\bibfnamefont {J.}~\bibnamefont {Ba}},\
	}\bibfield  {title} {\bibinfo {title} {Adam: {A} method for stochastic
			optimization},\ }in\ \href@noop {} {\emph {\bibinfo {booktitle} {3rd
				International Conference on Learning Representations (ICLR)}}}\ (\bibinfo
	{year} {2015})\BibitemShut {NoStop}%
	\bibitem [{\citenamefont {Porezag}\ \emph {et~al.}(1995)\citenamefont
		{Porezag}, \citenamefont {Frauenheim}, \citenamefont {K{\"o}hler},
		\citenamefont {Seifert},\ and\ \citenamefont {Kaschner}}]{porezagdftb1995}%
	\BibitemOpen
	\bibfield  {author} {\bibinfo {author} {\bibfnamefont {D.}~\bibnamefont
			{Porezag}}, \bibinfo {author} {\bibfnamefont {T.}~\bibnamefont {Frauenheim}},
		\bibinfo {author} {\bibfnamefont {T.}~\bibnamefont {K{\"o}hler}}, \bibinfo
		{author} {\bibfnamefont {G.}~\bibnamefont {Seifert}},\ and\ \bibinfo {author}
		{\bibfnamefont {R.}~\bibnamefont {Kaschner}},\ }\bibfield  {title} {\bibinfo
		{title} {Construction of tight-binding-like potentials on the basis of
			density-functional theory: {{Application}} to carbon},\ }\href@noop {}
	{\bibfield  {journal} {\bibinfo  {journal} {Phys. Rev. B}\ }\textbf {\bibinfo
			{volume} {51}},\ \bibinfo {pages} {12947} (\bibinfo {year}
		{1995})}\BibitemShut {NoStop}%
	\bibitem [{\citenamefont {Elstner}\ \emph {et~al.}(1998)\citenamefont
		{Elstner}, \citenamefont {Porezag}, \citenamefont {Jungnickel}, \citenamefont
		{Elsner}, \citenamefont {Haugk}, \citenamefont {Frauenheim}, \citenamefont
		{Suhai},\ and\ \citenamefont {Seifert}}]{Elstnerdftb1998}%
	\BibitemOpen
	\bibfield  {author} {\bibinfo {author} {\bibfnamefont {M.}~\bibnamefont
			{Elstner}}, \bibinfo {author} {\bibfnamefont {D.}~\bibnamefont {Porezag}},
		\bibinfo {author} {\bibfnamefont {G.}~\bibnamefont {Jungnickel}}, \bibinfo
		{author} {\bibfnamefont {J.}~\bibnamefont {Elsner}}, \bibinfo {author}
		{\bibfnamefont {M.}~\bibnamefont {Haugk}}, \bibinfo {author} {\bibfnamefont
			{T.}~\bibnamefont {Frauenheim}}, \bibinfo {author} {\bibfnamefont
			{S.}~\bibnamefont {Suhai}},\ and\ \bibinfo {author} {\bibfnamefont
			{G.}~\bibnamefont {Seifert}},\ }\bibfield  {title} {\bibinfo {title}
		{Self-consistent-charge density-functional tight-binding method for
			simulations of complex materials properties},\ }\href@noop {} {\bibfield
		{journal} {\bibinfo  {journal} {Phys. Rev. B}\ }\textbf {\bibinfo {volume}
			{58}},\ \bibinfo {pages} {7260} (\bibinfo {year} {1998})}\BibitemShut
	{NoStop}%
	\bibitem [{\citenamefont {Heimann}\ \emph {et~al.}(1983)\citenamefont
		{Heimann}, \citenamefont {Kleiman},\ and\ \citenamefont
		{Salansky}}]{Heimann1983}%
	\BibitemOpen
	\bibfield  {author} {\bibinfo {author} {\bibfnamefont {R.~B.}\ \bibnamefont
			{Heimann}}, \bibinfo {author} {\bibfnamefont {J.}~\bibnamefont {Kleiman}},\
		and\ \bibinfo {author} {\bibfnamefont {N.~M.}\ \bibnamefont {Salansky}},\
	}\bibfield  {title} {\bibinfo {title} {A unified structural approach to
			linear carbon polytypes},\ }\href@noop {} {\bibfield  {journal} {\bibinfo
			{journal} {Nature}\ }\textbf {\bibinfo {volume} {306}},\ \bibinfo {pages}
		{164} (\bibinfo {year} {1983})}\BibitemShut {NoStop}%
	\bibitem [{\citenamefont {Jin}\ \emph {et~al.}(2009)\citenamefont {Jin},
		\citenamefont {Lan}, \citenamefont {Peng}, \citenamefont {Suenaga},\ and\
		\citenamefont {Iijima}}]{Chuanhong2009}%
	\BibitemOpen
	\bibfield  {author} {\bibinfo {author} {\bibfnamefont {C.}~\bibnamefont
			{Jin}}, \bibinfo {author} {\bibfnamefont {H.}~\bibnamefont {Lan}}, \bibinfo
		{author} {\bibfnamefont {L.}~\bibnamefont {Peng}}, \bibinfo {author}
		{\bibfnamefont {K.}~\bibnamefont {Suenaga}},\ and\ \bibinfo {author}
		{\bibfnamefont {S.}~\bibnamefont {Iijima}},\ }\bibfield  {title} {\bibinfo
		{title} {Deriving carbon atomic chains from graphene},\ }\href@noop {}
	{\bibfield  {journal} {\bibinfo  {journal} {Phys. Rev. Lett.}\ }\textbf
		{\bibinfo {volume} {102}},\ \bibinfo {pages} {205501} (\bibinfo {year}
		{2009})}\BibitemShut {NoStop}%
	\bibitem [{\citenamefont {Cretu}\ \emph {et~al.}(2013)\citenamefont {Cretu},
		\citenamefont {Botello-Mendez}, \citenamefont {Janowska}, \citenamefont
		{Pham-Huu}, \citenamefont {Charlier},\ and\ \citenamefont
		{Banhart}}]{Cretu2013}%
	\BibitemOpen
	\bibfield  {author} {\bibinfo {author} {\bibfnamefont {O.}~\bibnamefont
			{Cretu}}, \bibinfo {author} {\bibfnamefont {A.~R.}\ \bibnamefont
			{Botello-Mendez}}, \bibinfo {author} {\bibfnamefont {I.}~\bibnamefont
			{Janowska}}, \bibinfo {author} {\bibfnamefont {C.}~\bibnamefont {Pham-Huu}},
		\bibinfo {author} {\bibfnamefont {J.-C.}\ \bibnamefont {Charlier}},\ and\
		\bibinfo {author} {\bibfnamefont {F.}~\bibnamefont {Banhart}},\ }\bibfield
	{title} {\bibinfo {title} {Electrical transport measured in atomic carbon
			chains},\ }\href@noop {} {\bibfield  {journal} {\bibinfo  {journal} {Nano
				Lett.}\ }\textbf {\bibinfo {volume} {13}},\ \bibinfo {pages} {3487} (\bibinfo
		{year} {2013})}\BibitemShut {NoStop}%
	\bibitem [{\citenamefont {Kertesz}\ \emph {et~al.}(1978)\citenamefont
		{Kertesz}, \citenamefont {Koller},\ and\ \citenamefont
		{Azman}}]{Kertesz1978}%
	\BibitemOpen
	\bibfield  {author} {\bibinfo {author} {\bibfnamefont {M.}~\bibnamefont
			{Kertesz}}, \bibinfo {author} {\bibfnamefont {J.}~\bibnamefont {Koller}},\
		and\ \bibinfo {author} {\bibfnamefont {A.}~\bibnamefont {Azman}},\ }\bibfield
	{title} {\bibinfo {title} {Ab initio {Hartree–Fock} crystal orbital
			studies. {II}. {Energy} bands of an infinite carbon chain},\ }\href@noop {}
	{\bibfield  {journal} {\bibinfo  {journal} {J. Chem. Phys.}\ }\textbf
		{\bibinfo {volume} {68}},\ \bibinfo {pages} {2779} (\bibinfo {year}
		{1978})}\BibitemShut {NoStop}%
	\bibitem [{\citenamefont {Peierls}(1996)}]{Peierls1996}%
	\BibitemOpen
	\bibfield  {author} {\bibinfo {author} {\bibfnamefont {R.~E.}\ \bibnamefont
			{Peierls}},\ }\href@noop {} {\emph {\bibinfo {title} {Quantum Theory of
				Solids}}},\ International Series of Monographs on Physics\ (\bibinfo
	{publisher} {Clarendon Press},\ \bibinfo {year} {1996})\BibitemShut {NoStop}%
	\bibitem [{\citenamefont {Su}\ \emph {et~al.}(1979)\citenamefont {Su},
		\citenamefont {Schrieffer},\ and\ \citenamefont {Heeger}}]{Su79}%
	\BibitemOpen
	\bibfield  {author} {\bibinfo {author} {\bibfnamefont {W.~P.}\ \bibnamefont
			{Su}}, \bibinfo {author} {\bibfnamefont {J.~R.}\ \bibnamefont {Schrieffer}},\
		and\ \bibinfo {author} {\bibfnamefont {A.~J.}\ \bibnamefont {Heeger}},\
	}\bibfield  {title} {\bibinfo {title} {Solitons in polyacetylene},\
	}\href@noop {} {\bibfield  {journal} {\bibinfo  {journal} {Phys. Rev. Lett.}\
		}\textbf {\bibinfo {volume} {42}},\ \bibinfo {pages} {1698} (\bibinfo {year}
		{1979})}\BibitemShut {NoStop}%
	\bibitem [{\citenamefont {Kresse}\ and\ \citenamefont
		{Furthm\"uller}(1996)}]{Kresse1996}%
	\BibitemOpen
	\bibfield  {author} {\bibinfo {author} {\bibfnamefont {G.}~\bibnamefont
			{Kresse}}\ and\ \bibinfo {author} {\bibfnamefont {J.}~\bibnamefont
			{Furthm\"uller}},\ }\bibfield  {title} {\bibinfo {title} {Efficient iterative
			schemes for ab initio total-energy calculations using a plane-wave basis
			set},\ }\href@noop {} {\bibfield  {journal} {\bibinfo  {journal} {Phys. Rev.
				B}\ }\textbf {\bibinfo {volume} {54}},\ \bibinfo {pages} {11169} (\bibinfo
		{year} {1996})}\BibitemShut {NoStop}%
	\bibitem [{\citenamefont {Perdew}\ and\ \citenamefont
		{Zunger}(1981)}]{Perdew1981}%
	\BibitemOpen
	\bibfield  {author} {\bibinfo {author} {\bibfnamefont {J.~P.}\ \bibnamefont
			{Perdew}}\ and\ \bibinfo {author} {\bibfnamefont {A.}~\bibnamefont
			{Zunger}},\ }\bibfield  {title} {\bibinfo {title} {Self-interaction
			correction to density-functional approximations for many-electron systems},\
	}\href@noop {} {\bibfield  {journal} {\bibinfo  {journal} {Phys. Rev. B}\
		}\textbf {\bibinfo {volume} {23}},\ \bibinfo {pages} {5048} (\bibinfo {year}
		{1981})}\BibitemShut {NoStop}%
	\bibitem [{\citenamefont {Nosé}(1984)}]{Nose1984}%
	\BibitemOpen
	\bibfield  {author} {\bibinfo {author} {\bibfnamefont {S.}~\bibnamefont
			{Nosé}},\ }\bibfield  {title} {\bibinfo {title} {A molecular dynamics method
			for simulations in the canonical ensemble},\ }\href@noop {} {\bibfield
		{journal} {\bibinfo  {journal} {Mol. Phys.}\ }\textbf {\bibinfo {volume}
			{52}},\ \bibinfo {pages} {255} (\bibinfo {year} {1984})}\BibitemShut
	{NoStop}%
	\bibitem [{\citenamefont {Hoover}(1985)}]{Hoover1985}%
	\BibitemOpen
	\bibfield  {author} {\bibinfo {author} {\bibfnamefont {W.~G.}\ \bibnamefont
			{Hoover}},\ }\bibfield  {title} {\bibinfo {title} {Canonical dynamics:
			Equilibrium phase-space distributions},\ }\href@noop {} {\bibfield  {journal}
		{\bibinfo  {journal} {Phys. Rev. A}\ }\textbf {\bibinfo {volume} {31}},\
		\bibinfo {pages} {1695} (\bibinfo {year} {1985})}\BibitemShut {NoStop}%
	\bibitem [{\citenamefont {Kohn}\ and\ \citenamefont
		{Sham}(1965)}]{KohnSham1965}%
	\BibitemOpen
	\bibfield  {author} {\bibinfo {author} {\bibfnamefont {W.}~\bibnamefont
			{Kohn}}\ and\ \bibinfo {author} {\bibfnamefont {L.~J.}\ \bibnamefont
			{Sham}},\ }\bibfield  {title} {\bibinfo {title} {Self-consistent equations
			including exchange and correlation effects},\ }\href@noop {} {\bibfield
		{journal} {\bibinfo  {journal} {Phys. Rev.}\ }\textbf {\bibinfo {volume}
			{140}},\ \bibinfo {pages} {A1133} (\bibinfo {year} {1965})}\BibitemShut
	{NoStop}%
	\bibitem [{\citenamefont {{Tishby}}\ and\ \citenamefont
		{{Zaslavsky}}(2015)}]{Tishby_2015}%
	\BibitemOpen
	\bibfield  {author} {\bibinfo {author} {\bibfnamefont {N.}~\bibnamefont
			{{Tishby}}}\ and\ \bibinfo {author} {\bibfnamefont {N.}~\bibnamefont
			{{Zaslavsky}}},\ }\bibfield  {title} {\bibinfo {title} {Deep learning and the
			information bottleneck principle},\ }in\ \href@noop {} {\emph {\bibinfo
			{booktitle} {2015 IEEE Information Theory Workshop (ITW)}}}\ (\bibinfo {year}
	{2015})\ pp.\ \bibinfo {pages} {1--5}\BibitemShut {NoStop}%
	\bibitem [{\citenamefont {Shwartz{-}Ziv}\ and\ \citenamefont
		{Tishby}(2017)}]{Tishby2017}%
	\BibitemOpen
	\bibfield  {author} {\bibinfo {author} {\bibfnamefont {R.}~\bibnamefont
			{Shwartz{-}Ziv}}\ and\ \bibinfo {author} {\bibfnamefont {N.}~\bibnamefont
			{Tishby}},\ }\bibfield  {title} {\bibinfo {title} {Opening the black box of
			deep neural networks via information},\ }\href@noop {} {\bibfield  {journal}
		{\bibinfo  {journal} {Computing Research Repository (CoRR)}\ }\textbf
		{\bibinfo {volume} {abs/1703.00810}} (\bibinfo {year} {2017})}\BibitemShut
	{NoStop}%
	\bibitem [{\citenamefont {Kullback}\ and\ \citenamefont
		{Leibler}(1951)}]{Kullback51}%
	\BibitemOpen
	\bibfield  {author} {\bibinfo {author} {\bibfnamefont {S.}~\bibnamefont
			{Kullback}}\ and\ \bibinfo {author} {\bibfnamefont {R.~A.}\ \bibnamefont
			{Leibler}},\ }\bibfield  {title} {\bibinfo {title} {On information and
			sufficiency},\ }\href@noop {} {\bibfield  {journal} {\bibinfo  {journal}
			{Ann. Math. Statist.}\ }\textbf {\bibinfo {volume} {22}},\ \bibinfo {pages}
		{79} (\bibinfo {year} {1951})}\BibitemShut {NoStop}%
	\bibitem [{Note1()}]{Note1}%
	\BibitemOpen
	\bibinfo {note} {Though no nonadiabatic effects on the ion dynamics is
		included in this approach.}\BibitemShut {Stop}%
	\bibitem [{\citenamefont {Plimpton}(1995)}]{lammps1995}%
	\BibitemOpen
	\bibfield  {author} {\bibinfo {author} {\bibfnamefont {S.}~\bibnamefont
			{Plimpton}},\ }\bibfield  {title} {\bibinfo {title} {Fast parallel algorithms
			for short-range molecular dynamics},\ }\href@noop {} {\bibfield  {journal}
		{\bibinfo  {journal} {J. Comput. Phys.}\ }\textbf {\bibinfo {volume} {117}},\
		\bibinfo {pages} {1} (\bibinfo {year} {1995})}\BibitemShut {NoStop}%
	\bibitem [{\citenamefont {Wang}\ \emph {et~al.}(2018)\citenamefont {Wang},
		\citenamefont {Zhang}, \citenamefont {Han},\ and\ \citenamefont
		{E}}]{deepmdkit2018}%
	\BibitemOpen
	\bibfield  {author} {\bibinfo {author} {\bibfnamefont {H.}~\bibnamefont
			{Wang}}, \bibinfo {author} {\bibfnamefont {L.}~\bibnamefont {Zhang}},
		\bibinfo {author} {\bibfnamefont {J.}~\bibnamefont {Han}},\ and\ \bibinfo
		{author} {\bibfnamefont {W.}~\bibnamefont {E}},\ }\bibfield  {title}
	{\bibinfo {title} {Deepmd-kit: A deep learning package for many-body
			potential energy representation and molecular dynamics},\ }\href@noop {}
	{\bibfield  {journal} {\bibinfo  {journal} {Comput. Phys. Commun.}\ }\textbf
		{\bibinfo {volume} {228}},\ \bibinfo {pages} {178} (\bibinfo {year}
		{2018})}\BibitemShut {NoStop}%
	\bibitem [{\citenamefont {Tuckerman}\ \emph {et~al.}(1993)\citenamefont
		{Tuckerman}, \citenamefont {Berne}, \citenamefont {Martyna},\ and\
		\citenamefont {Klein}}]{Tuckerman1993}%
	\BibitemOpen
	\bibfield  {author} {\bibinfo {author} {\bibfnamefont {M.~E.}\ \bibnamefont
			{Tuckerman}}, \bibinfo {author} {\bibfnamefont {B.~J.}\ \bibnamefont
			{Berne}}, \bibinfo {author} {\bibfnamefont {G.~J.}\ \bibnamefont {Martyna}},\
		and\ \bibinfo {author} {\bibfnamefont {M.~L.}\ \bibnamefont {Klein}},\
	}\bibfield  {title} {\bibinfo {title} {Efficient molecular dynamics and
			hybrid monte carlo algorithms for path integrals},\ }\href@noop {} {\bibfield
		{journal} {\bibinfo  {journal} {J. Chem. Phys.}\ }\textbf {\bibinfo {volume}
			{99}},\ \bibinfo {pages} {2796} (\bibinfo {year} {1993})}\BibitemShut
	{NoStop}%
	\bibitem [{\citenamefont {Nghiem}\ and\ \citenamefont
		{Costi}(2017)}]{Nghiem17}%
	\BibitemOpen
	\bibfield  {author} {\bibinfo {author} {\bibfnamefont {H.~T.~M.}\
			\bibnamefont {Nghiem}}\ and\ \bibinfo {author} {\bibfnamefont {T.~A.}\
			\bibnamefont {Costi}},\ }\bibfield  {title} {\bibinfo {title} {Time evolution
			of the kondo resonance in response to a quench},\ }\href@noop {} {\bibfield
		{journal} {\bibinfo  {journal} {Phys. Rev. Lett.}\ }\textbf {\bibinfo
			{volume} {119}},\ \bibinfo {pages} {156601} (\bibinfo {year}
		{2017})}\BibitemShut {NoStop}%
	\bibitem [{\citenamefont {Mahan}(2013)}]{mahan2013}%
	\BibitemOpen
	\bibfield  {author} {\bibinfo {author} {\bibfnamefont {G.~D.}\ \bibnamefont
			{Mahan}},\ }\href@noop {} {\emph {\bibinfo {title} {Many-particle physics}}}\
	(\bibinfo  {publisher} {Springer},\ \bibinfo {year} {2013})\BibitemShut
	{NoStop}%
\end{thebibliography}
%

\clearpage

\appendix

\renewcommand{\thefigure}{A-\arabic{figure}}
\setcounter{figure}{0}

\section{Soliton pair energy landscape}
\begin{figure}[htbp]
	\begin{centering}
		\includegraphics[width=70 mm]{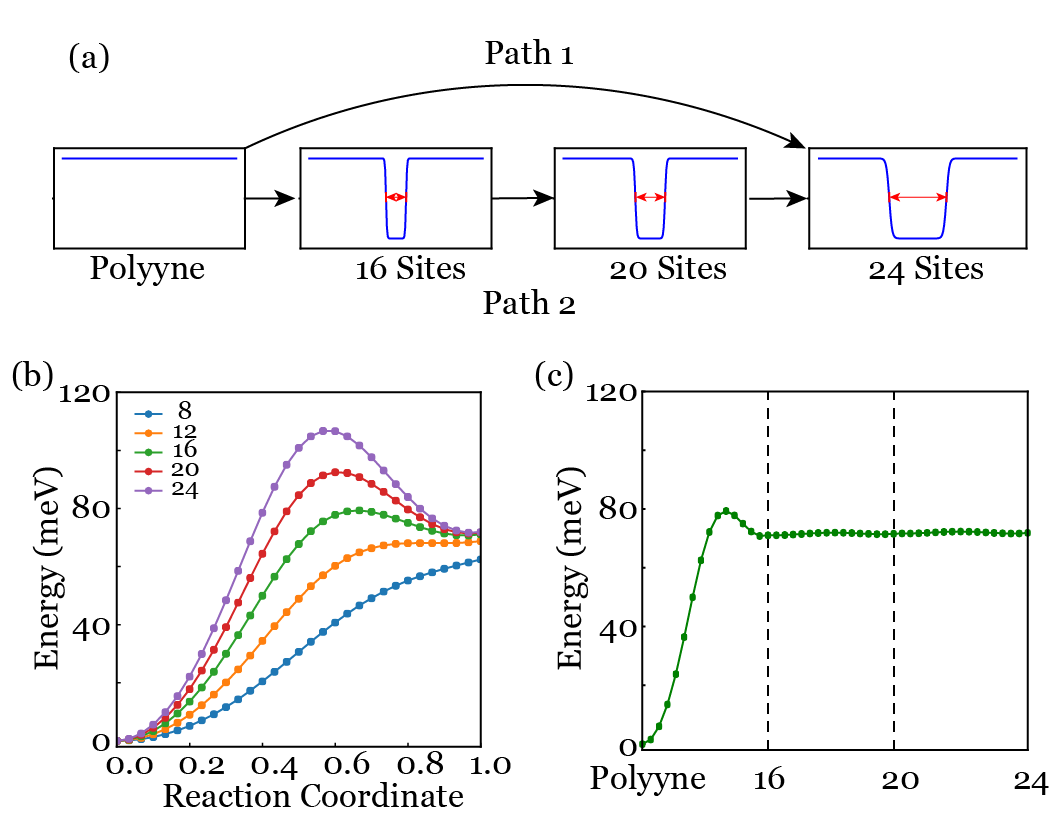}
	\end{centering}
	\protect\caption{Energy barrier for the transition between polyyne and soliton structure. (a) Direct (path 1) and indirect (path 2) transition paths from a polyyne to a soliton pair separated by 24 sites. (b) Direct transition barrier for the transition between polyyne structure and soliton structure with different separation between soliton and anti-soliton.(c) Indirect transition barrier for polyyne to a soliton pair as shown by the path 2 in (a).}
	\label{fig:s1}
\end{figure}

As illustrated in Fig.\ref{fig:s1}(a), there are two  pathways to a soliton-antisoliton pair in a polyyne, namely, the direct (path 1) and indirect (path 2) transitions. Fig.\ref{fig:s1}(b) shows the direct transition barriers from polyyne to the soliton structures with different separations between soliton-antisoliton pairs. The higher transition barrier evidently increases consistently with the spatial separation.  Therefore, a reasonable transition pathway should be a indirect path, whose potential profile is shown in Fig.\ref{fig:s1}(c). A soliton pair with small separation is generated from polyyne structure with the energy barrier $\sim 80$ meV, with ensuing nearly free parting of the incipient solitons to larger separations. 

Fig.\ref{fig:s1}(c) also indicates that annihilation for soliton-antisoliton pair is much easier than formation. Therefore, at low temperatures, classical MD can only sample the ground states polyyne structure. In path-integral MD, due to the quantum tunneling, the soliton phase is more frequently seen even at low temperatures.

\section{Optical conductivity calculation}

We describe how the optical conductivity\cite{mahan2013}
\begin{equation}
	\sigma_{\alpha\beta}(\omega) = 
	\frac{\ii}{\omega}\left[\int_{-\infty}^{\infty} dt e^{\ii \omega t} 
	G^R_{\alpha\beta}\left( t \right)
	+ \frac{n_0 e^2}{m} \delta_{\alpha\beta}\right]
\end{equation}
is calculated from TBworks-predicted tight-binding Hamiltonian.
Here, the retarded current-current correlation function is defined as
\begin{equation}
	G^R_{\alpha\beta}\left( t \right)=-\frac{\ii}{\hbar V} \Theta(t)\left\langle \left[j_{\alpha}(t),j_{\beta}(0)\right]\right\rangle.
\end{equation}

For a non-interacting system the current-current correlation function can be factorized easily by Wick's theorem, upon spectral decomposition
\begin{equation}
	G^R_{\alpha\beta}\left( \omega \right)=-\frac{\ii e^2} {\hbar V} \sum_{mn} \frac{f_m-f_n}{\varepsilon_m-\varepsilon_n+\omega+\ii 0^+}v_{mn}^\alpha v_{nm}^\beta. 
	\label{eq:corr}
\end{equation}
in which $f_n=(e^{\beta\left(\varepsilon_{n}-\mu\right)}+1)^{-1}$ is the Fermion-Dirac distribution function, and $\bm{v}_{mn}$ is the velocity matrix element
$
	\bm{v}_{mn} = \langle m |\vec v\,|n\rangle.
$

For a band structure system, we adopt a representation in which the first-quantized Hamiltonian is written
\begin{equation}
	H_{\vec k} = e^{\ii \vec k\cdot \vec r} He^{-\ii \vec k\cdot \vec r}.
\end{equation}
According to the Bloch theorem, an eigenstate of $H$ can be written as $\left|\psi_{n\vec k}\right\rangle =e^{i\bm{k}\cdot\bm{r}}\left|u_{n \vec k}\right\rangle$. In this representation, the velocity matrix element is calculated as 
\begin{equation}
	\vec v_{mn}(\vec k)  =\left\langle u_{n\vec k}\left| \hat {\vec{v}}(
	\vec k) \right|u_{m\vec k}\right\rangle,
\end{equation}
where the velocity operator is 
\begin{equation}
	\hat {\vec v}(\vec k) = \frac{1}{\hbar}\frac{\partial H_{\vec k}}{\partial \vec k}.
\end{equation}

From a set of localized Wannier functions corresponding to the tight-binding model,
$
	w_{a}\left(\bm{r}-\bm{R}\right)=\left\langle \bm{r}\left|a\bm{R}\right.\right\rangle,
$
the Bloch basis is formed,
$
	\left|\phi_{a\bm{k}}\right\rangle =\frac{1}{\sqrt{N}}\sum_{\bm{R}}e^{\ii \bm{k}\cdot\bm{R}}\left|a\bm{R}\right\rangle.
$
The Bloch Hamiltonian is
\begin{equation}
H_{ab}(\bm{k})  =\left\langle \phi_{a\bm{k}}\left|H\right|\phi_{b\bm{k}}\right\rangle 
 =\sum_{R}e^{i\bm{k}\cdot\left(\bm{R}\right)}H_{ab}(\vec R).
\end{equation}
Diagonalization of $H_{ab}(\bm{k})$ leads to bands $\varepsilon_{n\vec k}$
\begin{equation}
 U(\bm{k})^{\dagger}H (\vec k)U(\bm{k})=\operatorname {diag}(\varepsilon_{n\vec k}).
\end{equation}
where $U(\bm{k})$ is a unitary matrix composed of eigenvectors as columns.
Therefore, the velocity computed as
\begin{equation}
	\bm{v}\left(\bm{k}\right)= U(\bm{k})^{\dagger}\hat {\vec v} (\vec k)U(\bm{k})
	\label{eq:vel}.
\end{equation}
Finally, the current-current correlation function can be evaluated using Eq. (\ref{eq:corr}) together with Eq. (\ref{eq:vel}).


\section{Determination of system size for training data}
In main text, we present the TBworks model trained by labeled AIMD data with system size $N=32$. There is a natural lower bound for the simulation box size. We have obtained the localized Wannier functions for a perfect polycumulene crystal, this is shown in Fig.\ref{fig:s2} below. The Wannier function is plotted using 0.1\% the amplitude of that at the Wannier center as a cutoff, which is located at the 8$^{\rm{th}}$ nearest neighbor(NN). Thus, a minimal simulation box size is N = 17. We chose N = 32 for our training and the trained TBworks model predicts exact the tight-binding Hamiltonians as shown in main text.
\begin{figure}[hp]
	\begin{centering}
		\includegraphics[width=70 mm]{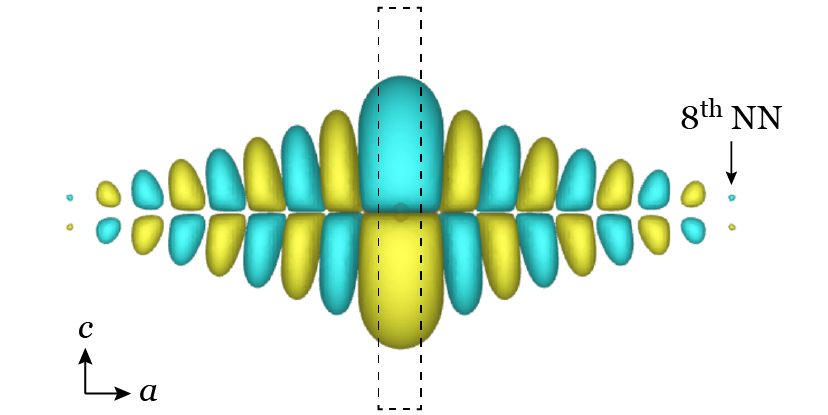}
	\end{centering}
	\protect\caption{The Wannier function for a perfect polycumulene crystal ploted using 0.1\% the amplitude of that at the Wannier center as a cutoff. Different color refers to the positive and negative value of the Wannier function.}
	\label{fig:s2}
\end{figure}

\section{Performance of TBworks trained by AIMD data with $N=48$}
\begin{figure}[htbp]
	\begin{centering}
		\includegraphics[width=70 mm]{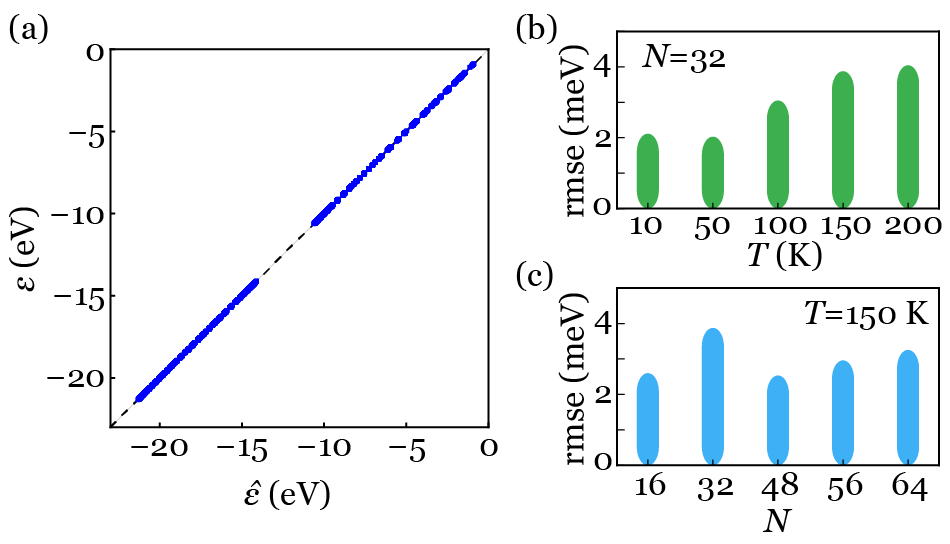}
	\end{centering}
	\protect\caption{Testing errors for the neural network trained by data set with $N= 48$, $T=150$~K.  (a) TBworks predicted vs \textit{ab initio} eigenvalues for unseen snapshots with different simulation box sizes and temperatures, $T=10, 50, 100, 150, 200$ K and $N=16, 32, 48, 56, 64$. (b) Testing error for unseen snapshots with a simulation box size N = 32 at different temperatures (c) Testing error for unseen snapshots with a temperatures T = 150K at different size.}
	\label{fig:s3}
\end{figure}

The performance of TBworks obtained by training with $N = 48$ data is similar as that training with $N = 32$ shown in main text. The performance of the model trained by $N=48$ data, shown in Fig.\ref{fig:s3}, shows that the neural network can predict highly accurate eigenvalues for unseen structures. In Fig.~\ref{fig:s3}(a), the TBworks (trained by $N=48$ data) predicted against \textit{ab initio} spectra for unseen snapshots are plotted, where samples are drawn from AIMD with multiple temperatures and a range of simulation box sizes as is done in Fig.~\ref{fig:2}(c).The testing error of these predictions is $\sim 3$~meV and the coefficient of determination $R^2= 0.9999997$. Fig.~\ref{fig:s3}(b)(c) also show that the TBworks trained at $N=48$ can also predict precisely the electronic spectra from multiple temperatures and simulation box sizes. Therefore, we conclude that the TBworks trained at  $N = 48$ can also offers a high fidelity representation of the \textit{ab initio} electronic structure and  is also  highly transferable to a range of temperatures and system sizes, same as the one trained at $N = 32$.

\clearpage
\end{document}